\documentclass[prb,aps,floatfix,twocolumn,amsmath,showpacs,groupedaddress,eqsecnum]
   {revtex4}
\usepackage{times}
\usepackage{graphicx}
\usepackage{amssymb}
\usepackage{verbatim}
\usepackage{psfrag}
\usepackage{color}
\usepackage[T1]{fontenc}
\usepackage{dcolumn}  

\def\CLH#1 {}
\newcommand{\LATER}[1] {}
\newcommand{\SOON}[1] {}
\def\TODO#1 {}
\def\TODOWC#1 {}
\def\DONEWC#1 {}
\def\NOTE#1 {}
\def\SAVE#1 {{}}
\def\OMIT#1 {{}}

\newcommand{\supcell}[3]{ $#1 \times #2 \times #3$ }
\newcommand{\ul}[1]{ \underline{#1} }
\newcommand{\ulO}{{\ul{\Omega}}}
\newcommand{\ulo}{{\ul{\omega}}}
\newcommand{\ulE}{{\ul{E}}}
\newcommand{\ulV}{{\ul{V}}}
\newcommand{\ulP}{{\ul{P}}}
\newcommand{\ulg}{{\ul{g}}}
\newcommand{\ulsigma}{{\ul{\sigma}}}
\newcommand{\VC}{V}   
\newcommand{\UC}{U}   

\newcommand{\HH}{\mathcal{H}}
\newcommand{\Ncell}{{N_{\rm cell}}}
\newcommand{\rr}{\vec{r}}
\newcommand{\dir}[3]{{#2} {#1} _{#3}}

\newcommand{\tet}{\hat{t}}

\makeatletter
\RequirePackage{keyval}
\define@key{placeholderkeyval}{width}{%
  \def\placeholderwidth{#1}}
\define@key{placeholderkeyval}{height}{%
  \def\placeholderheight{#1}}
\define@key{placeholderkeyval}{scale}{%
  \def\placeholderscale{#1}}
\newcommand{\donotyetincludegraphics}[2][width=0.8\columnwidth,height=0.5\columnwidth]{%
   \def\placeholderwidth{0.8\columnwidth}%
   \def\placeholderheight{0.5\columnwidth}%
   \setkeys{placeholderkeyval}{#1}%
   \fbox{\vbox to \placeholderheight{\hbox to \placeholderwidth{\hfill{\texttt{#2}}\hfill}}} \endgroup }
\makeatother

\begin{document}

\title{Orientational interaction and ordering of $Cd_4$ tetrahedra 
in a quasicrystal approximant}

\author{Woosong Choi}
\email{wc274@cornell.edu}
\author{Christopher L. Henley}
\email{clh@ccmr.cornell.edu}
\affiliation{Laboratory of Atomic and Solid State Physics (LASSP),
Clark Hall, Cornell University, Ithaca, New York 14853-2501, USA}
\author{Marek Mihalkovi\v{c}}
\affiliation{Slovak Academy of Sciences}

\begin{abstract}
We model the quasicrystal-related structure CaCd$_6$, 
a bcc packing of icosahedral clusters containing
tetrahedra which undergo orientational orderings
at T<100 K.  We use general schemes to evaluate an
effective Hamltonian for inter-tetrahedron orientations,
based on all-atom relaxations, either in terms 
of discrete cluster orientations, or of continuous
rotation angles.  The effective Hamiltonian is used
in Monte Carlo simulations to find the (complex) ground state
ordering pattern as a function of pressure. 
A preliminary investigation of thermal transitions found
(in part of the pressure range) two different first-order
transitions occurring below 100 K.
\end{abstract}
\maketitle

\section{Introduction}
\label{sec:intro}

Icosahedral $i$-CaCd (and isostructural alloys such as $i$-CdYb) are
the only known {\it binary} quasicrystals 
that have stable long-range-order.~\cite{Tsai00,Guo00,Ta01b}.
Furthermore, their atomic structures represent a third family
among quasicrystals, distinct from the previously known families
of aluminum-transition metal and of Frank-Kasper packing (though including
features of each).
The structure contains large
icosahedral ``Tsai'' clusters with {\it tetrahedra} of Cd atoms at their
centers, which obviously breaks the cluster's symmetry.  
is built from from icosahedral ``Tsai'' clusters consisting of
several concentric shells;
the outer shells have icosahedral symmetry, but the innermost
one is a Cd$_4$ tetrahedron which 
can relatively easily rotate to
different orientations. 
This obviously breaks the cluster's symmetry.  

This paper is concerned with the tetrahedra in CaCd$_6$,  
a bcc packing of the same Tsai clusters;
an equivalent phase is stable in many other systems that
form binary quasicrystals (e.g. Cd-Y, Cd-Eu, etc).
Periodic structures having a unit cell like a fragment of
a quasicrystal phase are called ``approximants'';  
CaCd$_6$ is the simplest approximant of the $i$-CaCd quasicrystal
and others are known in nature.
\OMIT{ Large approximants are useful
in simulations as a systematic and optimal way to approximate
the aperiodic quasicrystal using periodic boundary conditions.}

Diffraction on CaCd$_6$-type approximants has revealed various
order/disorder transitions, which are ascribed to orientational
ordering of the clusters.
Thus, as a function of pressure (up to 5 GPa), 
Cd$_6$Yb has a complex phase diagram with six phases \cite{watanuki2006}.
However, the exact orientations have not yet been determined 
experimentally or explained theoretically for any of these phases.
In this paper, we compute a comprehensive set of interactions
which, we hope, will predict the orientational ordering patterns 
and transition temperatures, and may serve as 
a starting point to address the role of orientations in
stabilizing the quasicrystal phase.


The first modeling of tetrahedron energies
in CaCd$_6$ ~\cite{No08} 
used ab-initio energies and started
by {\it assuming} the experimentally refined sites~\cite{gomez2003}.
Thus it was a sort of energy-guided fit, 
resolving the correlations in partially-occupied sites
found from diffraction.

Later, Brommer {\it et al}
performed a multiscale analysis
to understand the interaction and show the transition behavior:
first building inter-atomic potentials,~\cite{brommer2006} 
then modeling the nearest-neighbor interactions with 
42 parameters (minus 6 constraints)~\cite{brommer2007},
and finally performing some Monte Carlo simulations with this 
effective cluster Hamiltonian~\cite{brommer2007}.
Our paper should be considered a followup of this work.

\SAVE{
The inner shell has eight coarse sites forming a cube,
with two ways to fill 4/8 so as to make a tetrahedron
\cite{Pa71}.  Then, each tetrahedron corner site splits into 
three, where any one of each three is filled: thus
there are $2\cdot 3^4 = 162$ discrete possibilities.
Of the nine configuration types inequivalent by
cubic symmetry, the best was taken,  
and then a comparison was made of all options for
their relative orientation, within the 2-cluster
cubic cell.}


In this paper, we build a systematic method find the effective
interactions of tetrahedra (or similar inner clusters in
other materials) from numerical relaxations.
We give a natural way to eliminate arbitrary redundant freedoms
in the interaction, so as to ensure the physical relevance of 
the fitted parameters in our model Hamiltonian, 
leading to a better view of the interactions 
(Section~\ref{sec:redundancies}).
A singular-value decomposition is used to identify the
dominant contributions in the potential (and, in
principle, to reduce the number of terms needed to 
represent it).
We also show (Sec.~\ref{sec:results-continuous}) 
how to extend the same framework from 
a discrete set of orientations to the whole orientation space, 
and try to infer the functional form of the cluster orientation 
interaction Hamiltonian.
Moreover, in Sec.~\ref{sec:ordering}
we use the effective Hamiltonian to find the lowest energy states 
in super cells not accessed by 
Brommer's original calculations~\cite{brommer-thesis}.
We explore other low energy structures that are found in 
larger super cell Monte Carlo relaxations. 

\section{Framework and methods}
\label{sec:framework}

In this section, we introduce the general concepts
and methods used in the rest of the paper,
specifically the nature of the inter-cluster effective
Hamiltonian (Sec.~\ref{sec:cluster-H}), the
microscopic calculation of relaxed energies 
(Sec.~\ref{sec:pot-relax}), and our scheme for
extracting and processing the effective 
Hamiltonian (Sec.~\ref{sec:find-H}).
Most of them are not specific to Tsai clusters
with tetrahedra, but would also work for other
kinds of reorientable interior clusters inside
icoahedral clusters, such as the pseudo-Mackay
icosahedron in AlIr or AlPdMn materials~\cite{MM-AlIr}.

Before specializing to the clusters, we will review
the constant structure that surrounds them.
In CaCd$_6$, it consists of a bcc packing of icosahedral Tsai clusters,
with lattice constant 15.7\AA.
\LATER{Check lattice const. Is it 15.83?}
Each Tsai cluster consists of the following concentric shells:
\begin{itemize}
\item[(1)] Zn$_4$ tetrahedron, radius 1.9\AA;
\item[(2)] Zn$_{20}$ dodecahedral cage, radius $\sim 4.2$\AA;
\item[(3)] Ca$_{12}$ icosahedron, radius 5.56\AA;
\item[(4)] Zn$_{30}$ icosidodecahedron, radius $\sim 6.4$\AA.
\end{itemize}
These clusters touch along the 3-fold direction, while there are a 
few more Zn atoms between clusters around the 2-fold direction.
(Alternately, these Zn atoms may be reckoned to belong to large triacontrahedra
of Zn on both vertices and midedges, which overlap along the 
3-fold inter-cluster linkage.)

\subsection{Cluster effective Hamiltonian}
\label{sec:cluster-H}

In this material, the
low-energy degrees of freedom
are the tetrahedron cluster orientations, 
represented by rotation matrices $\{\ulO_i \}$ relative to
some reference orientation, where $i=1...\Ncell$ as there
is one cluster in each of the $\Ncell$ primitive cells.
The positions of all other atoms are taken to relax, so as to 
accommodate the tetrahedra;~\cite{mm2011-ScZn}.
they may, indeed, have large
displacements, but these are dependent on $\{ \ulO_i \}$.

We define an effective cluster Hamiltonian $\HH(\{\ulO_i \})$
as the minimum energy taken over all possible configurations 
constrained to have that combination of orientations, 
allowing relaxations of the surrounding atoms as well as 
distortions of the tetrahedra. 
\footnote{
Our definition is intended strictly for $T=0$.
At finite temperature, the effective Hamlitonian should properly 
be defined as a free energy, including both energy and entropy
contributions, and sampling all configurations with the
given tetrahedron orientations.  That would obviously be
much harder to evaluate from simulations.}
This effective Hamiltonian breaks up into 
one-, two-, and many-cluster terms:
\begin{equation}
  \HH({\ulO}) = \HH_1 + \HH_2 +\HH_3 + \cdots
  \label{eq:TotalE}
\end{equation}
We will assume that the many-body interactions are negligible.
Of course, the terms have the full symmetry of the crystal structure
(minus the tetrahedra): for example, 
the one-body term is the same
for all clusters and is invariant under the point group
$m\bar{3}$ (=$T_h$).
\SAVE{MM says the point group is simply $m\bar{3}$.}.

There is a useful analogy between the cluster degree of 
freedom and a (classical) spin, a unit vector that
is specified by only two Euler angles, in contrast
to the $3\times 3$ rotation matrix $\ulO$ 
which requires three Euler angles.
The two-cluster interaction is analogous to
dipolar or exchange spin interactions,  while
the single-body potential $\UC(\ulO)$ is analogous 
to the single-spin spin anisotropies due to crystal
fields.  An even closer analogy is to the interacting,
rotatable CN dipoles in KBr$_{1-x}$(CN)$_x$~\cite{grannan1988}.

\subsubsection{Single-cluster terms and optimal orientations}
\label{sec:one-cluster}

The single-body terms are
\begin{equation}
  \HH_1 = \sum_i \UC(\ulO_i) 
\end{equation}
The single-cluster potential $\UC(\ulO)$ includes a large 
contribution with icosahedral symmetry, reflecting the strong steric 
interaction between the tetrahedron and the dodecahedral
Cd$_{20}$ cage around it.
Because those cage atoms sit practically at the hard-core
radius, it is not surprising if $\UC(\Omega)$ has some 
sharp and irregular-looking dependences on orientation.
The single-body term is expected to have a somewhat smaller contribution 
with cubic symmetry, indirectly due to the Tsai cluster's outer shell being
distorted by its surroundings.
\SAVE{More generally, the second contribution reflects
the  reduced symmetry of the local environment in the
cluster packing. For example, 
that shell has cubic symmetry in the 1/1 approximant,
but not in the 2/1 cubic approximant since that has  
a non-symmorphic space group $Pa\bar{3}$.}

These tetrahedra get from one orientation to another by a
quasi-rigid rotation: rigid in the sense that the topological
identity of the tetrahedron is maintained throughout, but 
in fact both the tetrahedron and its Cd$_{20}$ cage 
undergo strong distortions.  (Indeed, sterically the tetrahedron
cannot never change orientations unless there are cooperative
motions of the caging atoms.)
A previous study~\cite{mm2011-ScZn}
addressed the barriers and dynamics of a single
Zn$_4$ tetrahedron in the Zn$_6$Sc approximant
(isostructural with CaCd$_6$).
The present paper is concerned only with static properties.


\begin{figure}[h]
  \begin{center}
    \includegraphics[width=0.64\columnwidth]{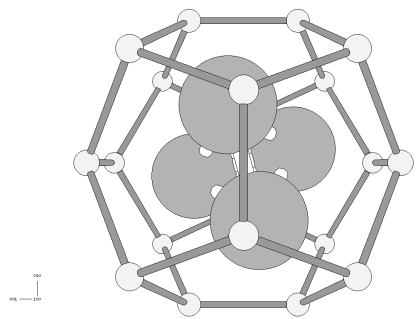}
  \end{center}
  \caption{Cd$_4$ tetrahedron in one of the twelve discrete optimal
configurations. For visibility, the tetrahedron atoms are shown by large
balls whereas those of the surrounding Cd$_{20}$ cage are shown by
small spheres.  Bonds are drawn among the cage's atoms to highlight
its dodecahedral shape.
The orientation shown is $\dir{X}{+}{r}$, relative to the axes indicated.
\SAVE{It is $\dir{X}{+}{r}$ when $x$ comes out of page, and $y$ axis
is to top. Confirmed WC 8/13/2012.}
\TODO{CLH+: I think it is more natural if $z$ is coming out
of the page, in which case this is $\dir{Z}{+}{r}$.  But
need to redraw the axes bigger.}
  }
  \label{fig:DiscreteOrientations}
\end{figure}

Simulations~\cite{brommer2007} found that 
tetrahedra in CaCd$_6$ tend to relax into one of 
twelve symmetry-related discrete orientations $\ulo_\mu$ (for $\mu=1\cdots12$),
which must be minima of the single-cluster potential $\UC(\Omega)$.  
The fact that similar orientations are seen, regardless
of how neighboring clusters are  oriented, indicates 
the single-body term is at least as strong as the two-cluster term.
However, certain discrete orientations are  unstable in the presence
of certain backgrounds of uniform surrounding orientations 
(see Sec.~\ref{sec:uniform-background} below): 
so the single-body term 
is not {\it much} stronger than the cluster interaction.

A tetrahedron in one of the ideal orientations is shown in 
Figure~\ref{fig:DiscreteOrientations}.
One of its twofold (actually $\bar 4$) symmetry axes is lined up with one of the cubic coordinate
axes; in the figure, this is the axis coming out of the paper.
Now, there is no $\bar 4$ symmetry in the icosahedron's point  group (or more 
pertinently, one in the $2/m3$ point group of the cluster center in cubic
CaCd$_6$).  Hence, the two ends of that twofold axis are inequivalent. 
(Indeed, Figure~\ref{fig:DiscreteOrientations}
shows the front two atoms are lined up under two atoms 
of the cage,
whereas the two back atoms are rotated 90$^\circ$.)
Hence, there are six possible directions for that orientation
axis which we label $\pm X$, $\pm Y$, $\pm Z$. 

It is not quite stable for the two ``front'' tetrahedron atoms
to line up directly under the two nearby cage atoms.
The cluster relaxes by rotating around the orientation axis
by an angle of approximately $\pm 15^\circ$ -- 
in either sense,
thus spontaneously breaking a symmetry and giving twelve 
symmetry-equivalent directions.
We indicate the rightwards (clockwise) or leftwards (counter-clockwise)
rotation, {\it as viewed from the tetrahedron center},
by a subscript $r$ or $l$, so our complete labels a form
are written $\dir{X}{+}{r}$, etc.
(Since orientations related by $r\leftrightarrow l$ are relatively
close, we anticipate they may have similar interactions.)

\LATER{Note that a different set of discrete orientations pertains
in ScZn$_6$ as studied in Ref.~\onlinecite{mm2011-ScZn}.
Also, make a corrected statement about the not-quite-correct
site list used in Ref.~\cite{brommer2007}.}

In earlier experiments, G\'{o}mez and Lidin studied the x-ray diffraction of 
MCd$_6$ approximants, where  M= Ca, Y,  or rare-earth.
They mapped out the continuous electron density inside
Tsai clusters, which they were able to interpret in terms of 
a host of split positions representing
tetrahedron orientational disorder, with
preferred orientations of a single kind related by symmetry~\cite{gomez2003}.
The apparent symmetry (see Ref.~\onlinecite{Pi06}, Fig. 3)
depended on the kind of large ion M, presumably reflecting the relative
importance of the icosahedral and cubic components in the single-body
term $\UC(\ulO)$ of the orientational Hamiltonian:
icosahedral for M=Tm and Lu, cubic for M=Tb, and intermediate for M=Ho or Er.
\SAVE{They did not try Sc.}
Their result for the case CaCd$_6$ agrees with the simulations
of Ref.~\onlinecite{brommer2007} as confirmed by our own:
the tetrahedra sit in an asymmetric orientation. 
\LATER{Check -- isn't the Gomez-Lidin rotation angle 45$^\circ$?
So there would be six basic symmetry related orientations.?}

\SAVE{The actual discrete positions seen
are due to the balance of
orientational energies in this particular -- cubic -- structure, 
so they have no transferability to other quasicrystal approximants.}

\subsubsection{Cluster pair terms}

The two-body term is written
\begin{equation}
  \HH_2 = \sum_{i,j} \VC_{ij}(\ulO_i, \ulO_j).
\end{equation}
The function $\VC_{ij}$ is translationally invariant,  
depending on sites $(i,j)$ only through the vector connecting 
them. It is expected to decay with separation. In this paper, the only
separations  included in the fit are the two kinds of nearest neighbors 
in the bcc lattice of cluster centers: the ``$b$'' linkage (separation vector 
equivalent to $[0,0,1]$) and the ``$c$'' type linkage (separation vector
equivalent to $[1/2,1/2,1/2]$).

We now comment on the possible atomic-scale origins of the 
cluster effective interaction; however the results of this 
paper do not depend on understanding that, nor will they resolve it.
A priori, one expects the cluster pair interaction 
has two kinds of contributions: mediated elastically,
via displacements of intervening atoms, or mediated by the
electron sea. The latter is expressed, within our framework,
by the EAM or pair potentials (see Sec.~\ref{sec:pot}, below)
and more specifically by the long range Friedel oscillations 
characteristic of pair potentials in a metal.
In contrast to the Al-TM and F-K classes of quasicrystal,
Friedel oscillations do not appear to be crucial
for the ``Tsai cluster'' class of quasicrystals \cite{brommer2006}, 
which suggests the elastically mediated interaction should be dominant.
Furthermore, if direct electronic interactions were dominant,
one would expect the interaction of two clusters separated 
by a $[1,0,0]$ type bond to be invariant under a simultaneous
rotation by $90^\circ$ around the bond direction, which is
not the case (see Sec.~\ref{sec:results-discrete}).
However, ab-initio calculations by Ishii and collaborators
ound a substantial cluster interaction when atom positions are not 
relaxed, demonstrating that the direct electronic interactions
are significant.  
Finally, we remark that in the isostructural compound ScZn$_6$,
the cluster-cluster interaction is much smaller~\cite{mm2011-ScZn}.
\LATER{Get citations to Ishii.}

Since the two-cluster term is mediated by a comparatively small
distortion of the outer shells of the Tsai cluster, and/or is a sum
of several potential terms, we anticipate that 
it is more smoothly behaved, and that the contributions from 
different neighbors will be additive.

\subsection{Interatomic potentials}
\label{sec:pot-relax}

In order to do our fitting, we must build a database of
relaxed energies coming from a lower, more exact level of
description.  
We defined the cluster Hamiltonian as the relaxed minimum 
energy of the system, having fixed the orientation
of every cluster.
This opens up three questions:
\begin{itemize}
\item[(1)] 
How do we define or compute the energy of an arbitrary 
atomic configuration?
(Sometimes these energies can be computed
directly from ab-initio relaxations, but here we needed
to use ``classical'' potentials.~\footnote{The ``classical''
potentials are of course by anchored fitting to electronic
calculations that are based, via the Local Density Approximation,
to the many-electron Schr\"odinger equation.  They are classical 
in the sense that they are deterministic functions 
of the atomic positions, having integrated out the electron
wavefunctions and neglected the ionic zero-point motions.})
\item[(2)]
considering that the tetrahedra are typically 
distorted, what is our precise definition of
cluster orientations? (This is required in
order to define the family of configurations
we are minimizing over.)
\item[(3)]
How do we implement this constrained minimization
reliably?
\end{itemize}
This section explains our answers to each question
including the important technicalities.

\subsubsection{Potentials}
\label{sec:pot}

Classical potentials are essential in various situations
when molecular dynamics or relaxation is required in
supercells containing many clusters. For the present
problem, we used the minimum possible supercell which
is $3\times 3\times 3$ or about 4000 atoms, which 
is too large for doing repetitive ab-initio
relaxations.

\SAVE{Another example where  we need a supercell is
molecular dynamics. MD using  pair potentials showed 
that a excess at low frequencies
in the experimental phonon density of states of ScZn$_6$
(it does not scale with $\omega^2$ as an acoustic mode should),
can be explained by disorder in the Zn$_4$ tetrahedra orientations.
[MM still unpublished].}

Most of our calculations are based on the embedded-atom method (EAM) 
potentials fitted by Brommer and G\"ahler~\cite{brommer2006}.
Their method of fitting is laborious and cannot be quickly repeated
for a new material.  As an alternative, we also tried the empirical 
oscillating pair (EOP) potentials~\cite{EOPP}, which can be 
rapidly computed for any composition, but are valid only
while the conduction electron concentration is held constant.
Comparing results from the two potentials, as done in
Sec~\ref{sec:results-pair-pot} below, may give a 
measure of the uncertainty of our conclusions, and/or a
measure of the reliability of plain pair potentials in
this system where their validity is less assured.

\begin{figure}[h!]
  \begin{center}
  \includegraphics[width=0.8\columnwidth]{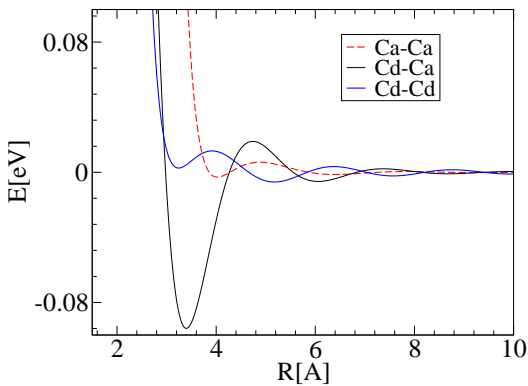}
  \end{center}
  \caption{Fitted ``empirical oscillating pair potentials''
(EOPP) for the Ca--Cd system.}
  \label{fig:EOPP-CaCd}
  \end{figure}

 The (EOPP) pair potentials use an six-parameter analytic form 
  which incorporates Friedel  oscillations~\cite{EOPP}.
  For this paper, they were fitted against a database of ab-initio 
  results with a total of 28 energy datapoints 
\TODO{MM 2a: (LATER) all of them (?)}
  taken from relaxed $T=0$K structures, plus a single snapshot from
  a high--temperature molecular dynamics simulation at 1000K 
  (of the cubic CaCd$_6$ structure) 
  that gave over 7000 force datapoints.
   Our database of relaxed samples included all known Ca--Cd binary
  compounds (CaCd$_2$ in both CeCu$_2$ and MgZn$_2$ structures; Ca$_3$Cd$_2$, 
  Ca$_2$Cd$_7$, and the B2 structure of CaCd;
  and versions of the CaCd$_6$ approximant
 with six different ways of placing two Cd$_4$
  tetrahedra in the cubic 1/1 cell.
  Furthermore, we added
  structures that we took from similar systems, such as Ca$_3$Zn, Ca$_2$Cu, CaZn$_3$, 
  CaCd$_{11}$ and finally the Frank-Kasper ``Bergman'' phase strucure 
  of AlMgZn.  In the final iteration, the fit was
  biased so as to give forces as accurate as possible. 
  \TODO{MM 2b: (LATER) biased, or weighted?}
  The results are shown in Figure~\ref{fig:EOPP-CaCd}

It should be noted that the Tsai class of quasicrystals
(and related alloys) is based on either Cd or Zn, both of which
are known in their elemental forms to have an anomalous
$c/a$ ratio in the hcp lattice.  Indeed, this might
be related to the extremely non-rigid  behavior of
the Zn$_{20}$ or Cd$_{20}$ dodecahedral cages in the Tsai
clusters, which is essential in allowing the inner
tetrahedra to rotate at all.  (See frames from
the finite-temperature MD simulation of ScZn$_6$,
Figure 1 of Ref.~\cite{mm2011-ScZn}).  We also know
that pure Zn is one of the few cases in which 
the EOP potentials more or less fail~\cite{EOPP}.
Consequently, it is somewhat surprising that we
find the EOP potentials succeed in CaCd$_6$, 
in that the cluster Hamiltonian fit is similar
to the result from EAM potentials (see Sec.~\ref{sec:results-pair-pot}).

\subsection{Implementing cluster orientations and constrained relaxation}
\label{sec:cluster+relax}

We need to establish an explicit practical mapping 
between atom positions and cluster orientations,
the degrees of freedom at the two levels of
description we want to relate.
First we lay out the atoms-to-orientation mapping,
and then its inverse, which actually means 
defining constraints for relaxation.

\subsubsection{Defining cluster orientations}
\label{sec:cluster-orientations}

It is relatively easy to define the orientation $\ulO$:
the four inner atoms always {\it do} form some kind of tetrahedron,
since it is sterically impossible for one of the atoms
to pass through the plane formed by the other three.

Let $\rr_l$ be the position of tetrahedron atom $l$
(for $l=1,...,4$), and define the center as 
$\rr_c\equiv (\rr_1+\rr_2+\rr_3+\rr_4)/4$; note that
$\rr_c$ can deviate from the center of the surrounding cage.
Also, define a regular reference tetrahedron by four unit
vectors $\{ \tet_l\}$  in tetrahedral directions.
(In the relaxation code, each tetrahedron's initial prescribed orientation 
is used as the reference.)
\SAVE{I.e., during the constraining step the (relative) 
orientation is set to identity (note from WC 9/11/12)}

Define a matrix $\ul{M}$ with components 
    \begin{equation}
       M_{\alpha\beta} \equiv 
       \frac{3}{4} \sum _{l=1}^4  (\rr_l-\rr_c)_\alpha (\tet_l)_\beta.
    \label{eq:M-build}
    \end{equation}
Now write the polar (=singular value) decomposition
    \begin{equation}
                 \ul{M} =\ulO_L \; \ul{M}_D \; \ulO_R
    \label{eq:polar}
    \end{equation}
where $\ulO_L$ and $\ulO_R$  are $3\times 3$ rotation matrices, 
and $\ul{M}_D$ is (positive) diagonal.
It is easy to check that if $\{ \rr_i \}$ is a regular tetrahedron,
then $\ul{M}_D$ is a multiple of the identity and the actual
tetrahedron is rotated, relative to  the reference tetrahedron,
by
    \begin{equation}
        \ulO=\ulO_L \ulO_R
    \label{eq:Omega-def}
    \end{equation}
For a general tetrahedron, we take Eq.~\eqref{eq:Omega-def} as our
{\it definition} of its orientation.  It can be shown that 
\eqref{eq:Omega-def} optimizes a measure of
agreement, $\sum_{l=1}^4 (\rr_l-\rr_c)\cdot \ulO \tet_l$.
If we had to do this for some other cluster with $n_a$ atoms, 
we simply need a different set of ideal vectors and the above
recipe still works, with the replacement $3/4 \to 3/n_a$ in
\eqref{eq:M-build}.

There is an alternative way to think of orientation extraction, 
which is specific to the (present) case that the cluster has 
exactly four atoms, not in the same plane.
We can uniquely and exactly represent the actual coordinates as
    \begin{equation}
            \rr_l = \ul{M} \tet_l + \rr_c
    \end{equation}
since this is a set of $4(3)$ linear equations in 
$3^2 + 3$ unknowns (the components of $\ul{M}$ and $\rr_c$.
In materials science, such a matrix defining an affine transformation 
of the atoms is called the {\it deformation matrix}. Indeed,
for four atoms it gives the same result for $\ul{M}$ 
as \eqref{eq:M-build}.  As above, a polar decomposition 
\eqref{eq:polar} is performed to define $\ulO$.

\subsubsection{Constrained relaxation}
\label{sec:constrained-relax}

\TODO{CLH: review CG algorithm}
In each constrained relaxation iteration, we assume 
a starting configuration in which 
each cluster already has its prescribed orientation,
according to the definition based on polar decomposition.
Conjugate gradient directions are constructed in the standard
(unconstrained) fashion, but are then projected orthogonally
into the allowed subspace, and one-dimensional minimizations
are carried out along this projected direction.

We relax atom configurations, subject to the EAM or pair potentials,
using a nonlinear conjugate gradient algorithm
with Newton-Raphson and Fletcher-Reeves~\cite{CG}.
Each successive conjugate-gradient iteration consists
of a one-dimensional Newton-Raphson minimization (with up
to 10 iterations) along the next conjugate-gradient direction.  
The stopping criterion is that
the energy change per step is $\Delta E_{\rm step} 
< 10^{-4} {\rm eV}$. 
Typically, a few hundred conjugate-gradient iterations were
needed. 
\SAVE{The maximum number of conjugate-gradient iterations
allowed is $N_{\rm max step}=1000$ (WC note in 9/11/12 draft).}

The question is how to constrain
all tetrahedron orientations $\{ \ulO_i \}$,
while relaxing  all atom coordinates.
This amounts to $3\Ncell$ nonlinear constraints,
defined implicitly by \eqref{eq:polar} and \eqref{eq:Omega-def}.
The basic approach is to linearize these 
constraints, defining a linear subspace within the 
\SAVE{$3N$-dimensional}
manifold of all atomic coordinates.

\SAVE{To start, note that in CaCd$_6$ the distortions of tetrahedra 
(as well as the displacements of $\rr_c$) are sufficiently 
great that it would be highly inaccurate to simply constrain 
the clusters to their unrelaxed positions while
relaxing all the other atoms.}

\SAVE{In an older version of code, using MM's relaxation code,
first {\it unconstrained} displacements are calculated by the
conjugate algorithm for all atoms.
The linearized conditions on the infinitesimal
displacements of the four atoms (for preserving their orientation)
were also computed.  Then the displacements of the tetrahedron atoms 
were projected onto that linear subspace of permitted changes.
But this was slightly changed in the final method using WC's
relaxation code. (WC node in 9/11/12 draft)}

\SAVE{WC note:For each relaxation step,
\begin{itemize}
  \item Conjugate gradient direction is determined
  \item Linear constraint applied to the direction
  \item Newton-Raphson minimization along the direction
  \item (Every $n$-times) Nonlinear reorienting performed
\end{itemize}}

\SAVE{CLH notes a warning.
Unfortunately, we cannot use the polar decomposition
idea to define an equivalence relation between arbitrary tetrahedra,
of having the same orientation: if
   \begin{equation}
{\rr''}_l = {\ulO''}_L {\ul{M}_D}'' {{\ulO''}_R}^{-1} {\rr'}_l
   \end{equation}
and
   \begin{equation}
       {\rr'}_l = {\ulO'}_L {\ul{M}_D}' {{\ulO'}_R}^{-1} \rr_l,
   \end{equation}
this implies
   \begin{equation}
{\rr''}_l = {\ulO''}_L {\ul{M}_D}'' {{\ulO''}_R}^{-1} {\ulO'}_L {\ul{M}_D}'
{\ulO'_R}^{-1} \rr'_l, 
   \end{equation}
and {\it not}
   \begin{equation}
      {\rr''}_l = \ulO_L \ul{M}_D {\ulO_R}^{-1}  {\rr'}_l.
   \end{equation}
}    

Since the actual constraint is nonlinear, after some iterations 
the configuration would not exactly satisfy it.  Therefore,
every $\sim 10$ iterations we perform a nonlinear projection to
reassert the orientation constraint.  
\TODO{CLH: how many iterations -- did Woosong send this number?  Ask WC.}
Namely, the actual
relaxed positions $\{ \rr'_l\}$ of the atoms in a tetrahedron 
are related to the ideal rotated positions $\{ \ulO \tet_l\}$ 
by a deformation matrix $\ul{M}'$, then polar decomposed as 
$\ul{M}'=\ulO_L \ul{M}_D \ulO_R$ as in \eqref{eq:polar}.
We then replace $\ul{M'}\to \ul{M}_D$

Specifically, let $\ul{M}'$ be a possible additional 
infinitesimal deformation of the four atom positions 
$\{ \rr'_l \}$ after a step, 
{\it relative to the previous positions}  $\{\rr_l\}$.
   \begin{equation}
       \rr^\prime-\rr_c =\ul{M}'(\rr-\rr_c), 
   \end{equation}
with $\ul{M}' = \ul{I}+\ul{m}'$ with $\ul{m}'$ small. 
Then the condition that $\ul{M}'$ contains no
rotation is that $\ul{m}'$ is symmetric, which 
implicitly defines a set of linear conditions on 
$\rr^\prime-\rr_c$.

\SAVE{In principle, we could relax without this linear constraint, 
and apply the nonlinear constraint at the end, and iterate till
convergence is achieved.  However if the linear constraint is not enforced, 
then relaxation competes with reorientation and screws up convergence. 
Something related happened with M.M.'s relaxation code, which is NOT
a simple Newton-Raphson type, and gave slower convergence
than this constrained CG.}

\SAVE{The orientation constraint is applied to ALL clusters,
not just the two ones used for the fitting.}

\subsubsection{Comparison to unconstrained relaxation method}

We want to constrast our constrained relaxation
approach with the simpler alternative approach
used in Ref.~\onlinecite{brommer2007}: plain
{\it unconstrained} relaxation starting with 
prescribed initial orientations (preferably,
optimal ones).
One difference is that our constrained relaxation 
allows construction of a full Hamiltonian as
a {\it continuous} function of orientations,
as carried out below in Section{sec:results-continuous},
which allows simulating the orientation fluctuations 
found at $T>0$ or mapping out the energy barriers for 
a cluster to reorient.

Using unconstrained relaxation, one accesses only the discrete set of orientations
that are local minima. Usually, one implicitly depends on the  assumption of
a one-to-one correspondence between the combinations of initial
orientations and final ones.  In reality, it may happen -- and did so
in our study, in a few instances -- that certain combinations of the
discrete orientations in neighboring clusters are {\it not} locally
stable: they relax into some other combination of orientations.
A final, technical drawback to unconstrained relaxation
is that even if that one-to-one correspondence exists,
the actual ground state orientations deviate from the
single-body optimum ones.  

\SAVE{It could happen that the actual two-body
interaction has a simple form, but the additional energy contribution 
reflecting its competition with the single-body terms may complicate
the form of the effective interaction.
To see how this might
happen, the reader is invited to consider the simpler analog of a
set of spin variabls ${\bf s}_i$ with interactions 
${\bf s}_i\cdot {\bf s}_j$
and a strong single-spin cubic anisotropy $-(s_{ix}^4+s_{iy}^4+s_{iz}^4)$.}

\subsection{Extracting the cluster Hamiltonian}
\label{sec:find-H}

It is assumed we start by choosing a discrete list of
representative possible orientations $\{ \ulo_\alpha, \alpha=1...m\}$.
In this work, these are either the twelve optimal
orientations of the single-body terms (to obtain the
discrete Hamiltonian in Sec.~\ref{sec:results-discrete}) 
or else a finer-spaced grid that is meant to sample
a continuous range of orientations (in Sec.~\ref{sec:results-continuous}).
Except where noted, our set of orientations $\{\ulo_\alpha\}$
always has the full point symmetry of cluster site.

A dataset is then constructed of the relaxed energies 
$E_{\alpha\beta}$ for orientation $\ulo_\alpha$ in cluster
site $i$ and $\ulo_\beta$ in cluster site $j$, for all 
combinations of the two, while the other (``background'')
clusters are held fixed.  In CaCd$_6$, we consider 
(mainly) two kinds of pairs, which are the nearest-neighbor
sites of the bcc lattice:  those separated by $\langle 100 \rangle$
vectors and by $\langle 1/2 1/2 1/2 \rangle$ vectors.
(We sometimes call these, respectively, ``$b$'' and ``$c$'' linkages,
based on their role in the canonical-cell tiling~\cite{Hen91CCT}.)

Our aim in fitting is to convert the array
$E_{\alpha\beta}$ to an energy function in 
the form of the cluster Hamiltonian~\eqref{eq:TotalE},
with well-defined and compact formulas for the single-body term 
$U(\Omega)$ and pair term $V_{ij}(\Omega,\Omega')$.


\subsubsection{Uniform background approach to isolate pairs}
\label{sec:uniform-background}

In order to properly fit the pair interaction of two clusters,
they must be put in a sufficiently large supercell that they have 
only one significant interaction with each other (no interactions with an image
of the neighbor through periodic boundary conditions in a 
different direction).  
\SAVE{Ideally, this should also be true for 
their interactions with ``environment'' clusters.}
When the cluster pair is 
related by the $b$ type vector [0,0,1], that demands a supercell of
at least $3\times 3 \times 4$ basic cells (i.e. 72 clusters).

It is not feasible to exhaustively enumerate all possible 
combinations of cluster orientations in the supercell
and relax every configuration; our fits must be based 
on a subset of orientations.  We will describe two 
different ways to choose this subset: exhaustive enumeration
of a pair in a fixed background, which we used in this
project, or random configurations of the whole system,
used by Brommer {\it et al}~\cite{brommer2007}.

Each supercell configuration is specified by three orientations:
$\ulO_i, \ulO_j$, the orientations of the two clusters
whose interaction we want, and the background orientation 
$\ulO_{bg}$, which is taken by all other clusters.
While keeping $\ulO_{bg}$ fixed, we enumerate
all $m^2$ possible combinations of $(\ulO_i,\ulO_j)$ and
(as described shortly below) extract a fit $V_{\alpha \beta}$.
Thus, we get $m$ independent fits, one for each background.
(Due to symmetries not all of them are independent.)
This provides some useful checks.

When our sampled orientations $\{\ulO_\alpha\}$ are just the
twelve optimal ones, which are symmetry equivalent, 
$U(\ulO_\alpha)$ has the same value for every one so
the correct result should be $U_\alpha=0$.  However, the
presence of a particular background breaks the symmetry;
in our fit, the pair interactions between cluster $i$ or $j$ and 
``background'' clusters all get absorbed into $U_\alpha$ term,
so it will be non-constant.  But if we average $U_\alpha$ over
all possible backgrounds, the symmetry should be restored.
After we complete the fit for $V_{\alpha\beta}$, we can 
predict the apparent single-body term due to the
background, and check if this is consistent with the 
apparent single-body terms that were actually fitted.
(Any disagreement suggests the importance of 
further-neighbor interactions with the background.)

In any case, provided we only have pair cluster interactions,
we should still obtain an identical fit for $V_{\alpha\beta}$
regardless of the background.  But actually, multi-body interactions 
including both of the selected clusters, and one or more of the 
background clusters, many contribute to the fitted $V_{\alpha\beta}$.
Thus, any dependence of $V_{\alpha\beta}$ on the background 
signals the presence of multi-body interactions.

\SAVE{Using a random but fixed background, and varying
only the two selected clusters, should be equally
valid but has not been tried.
Since a uniform background can cause the pair configuration
to be unstable, a random background might help with that.
On the other hand, some of the above mentioned checks would
not be possible.}

The uniform-background procedure relies on being able to 
generate any specified combination of orientations.
This might fail, if we use unconstrained relaxation,
because certain combinations of orientations might be
unstable and turn in a different direction.
In particular, we found that the background orientation 
$\ulO_{bg}$ is $\dir{X}{+}{r}$, 
a varied cluster with orientation $\dir{X}{-}{r}$ very often is unstable, 
reorienting to $\dir{X}{-}{l}$.
\SAVE{This might be due to the interaction between 
orientations $\dir{X}{+}{r}$ and $\dir{X}{-}{r}$ in the $x$ direction. 
[That is a background interaction,  since the interacting
clusters are related by (0,0,1)].}  
This problem is fixed by using a relaxation constrained to
a particular orientation (see Sec.~\ref{sec:constrained-relax}, above).
Alternatively, it may be evaded by using the random-whole-system 
approach instead.

The random-whole-system approach was used by 
Brommer {\it et al}~\cite{brommer2007} for CaCd$_6$.
(It was also used~\cite{Ri12} in Ir$_{23}$Al$_4$, a quasicrystal
approximant from the Al-transition metal class, which 
has a different kind of orientable inner shell consisting of
Al$_{10}$Ir in the pseudo-Mackay icosahedron cluster.)
Produce $10^3$--$10^4$ realizations of a supercell with a
random combination of orientations and relax each one;
if orientations are unstable, it is necessary to analyze the
final state to ascertain the actual index $\alpha$ for each
cluster.  In each configuration, we find the count $N^l_{\alpha\beta}$
of cluster linkages of type $l$ having the clusters in 
orientations $\ulO_\alpha$ and  $\ulO_\beta$, respectively,
Then we use a linear least-squares fit to express the total
energy as
   \begin{equation}
           E_{rm tot} = \sum_{l,\alpha,\beta} E^{l}_{\alpha\beta} 
           N^{l}_{\alpha\beta}.
   \end{equation}
Of course, symmetries are normally used so as to reduce the
number of parameters. 

\SAVE{The random sampling of configurations might conceivably
be taken as those visited in a finite-temperature molecular dynamics
or Monte Carlo simulation, or those found by relaxation
from random initial conditions.  The problem is that certain
neighbor pairs might never appear and therefore the fit would
be ill-defined (that is also a worry if some pairs 
are unstable to reorientation).}

We mention briefly a third way to construct a database, 
related in spirit to the uniform-background.  (It is analogous to
determining spin-spin interactions in a magnetic material
by diluting it with nonmagnetic ions that are chemically
equivalent to the magnetic ones, so that only isolated
pairs of the latter occur.)
We choose two clusters to vary through all combinations of
interactions, but take a background having {\it no} orientational interactions, 
by replacing every other tetrahedron by a single large atom that
takes up the same space; as it happens, the Ca atom is close to
the right size.  There is no averaging over backgrounds since there
is only one kind.

The single-atom replacement trick was successfully applied to
the isostructural compound ScZn$_6$, but only to study the
effects of the single-cluster term on the dynamics (the inter-cluster
interactions are, in any case, weak in that material).  
We experimented with this method for pair interactions in
CaCd$_6$, but these preliminary
investigations suggested that the systematic errors are too large 
for us to trust the results quantitatively.

\subsubsection{Symmetry properties of the discrete interaction matrix}
\label{sec:interaction-symmetries}

Consider the 
permutation symmetries of the interaction matrix $\ulV$ 
that follow from the space-group symmetry operations of the CaCd$_6$
framework, which affect $\ulV$ in two ways.
First, the rotation part of these operations permutes orientations
within the set $\{ \ulO_\alpha \}$ (except in the trivial case of pure translations).
Second, there are three interesting cases for the action on the
cluster centers: both may map to themselves, they may be swapped, or
one maps to itself and the other maps to a different cluster.
For each of these cases, it will be convenient to write $E_{\alpha\beta}$ and $V_{\alpha\beta}$ 
using matrix notation as arrays $\ulE$ and  $\ulV$,
which in general are {\it not} symmetric.
The action of a point group operation $g$ on cluster orientations
is written with an $m\times m$ {\it permutation matrix}, $\ulP_g$.
(Here we mean the matrix in which each row and column contains exactly one element 1, 
and the others are zero, so multiplication by this matrix induces
a permutation of the indices.)
\SAVE{(Evidently an orthogonal matrix.)}

First, for a lattice symmetry operation that keeps
both clusters fixed, invariance of the energy under
this symmetry is represented by
   \begin{equation}
           \ulP_g \; \ulV \; (\ulP_g)^T = \ulV.
   \label{eq:V-invariance}
   \end{equation}
This requires that the bond vector is invariant when $g$ acts at the 
midpoint.

Second, consider a rotational symmetry $h$ that exchanges the two clusters.
(In practice we need only consider inversion; all others are products
of inversion and a symmetry of the first kind.)
Now we get
   \begin{equation}
           \ulP_h \; \ulV \; (\ulP_h)^T = \ulV^T.
   \label{eq:V-swap}
   \end{equation}
\SAVE{Eq.~\eqref{eq:V-swap} implies that $\ulV$ is similar
to $\ulV^T$.}

Thirdly, consider a rotation $g$ that,
acting on site $i$ and maps $i\to i$ but sends $j\to k$.
We get
   \begin{equation}
           \ulP_g \; \ulV \; (\ulP_g)^T = \ulV'
   \label{eq:V-rotate-bond}
   \end{equation}
The use of \eqref{eq:V-rotate-bond} is that, having computed
the interaction matrix for a pair separated by (say) $[0,0,1]$,
we find the interaction matrix for symmetry related bond vectors
such as $[1 0 0]$.

\subsubsection{Redundant parameters: pair interaction resolution}
\label{sec:redundancies}
\label{sec:pair-int-resolve}

Whenever one fits an effective Hamiltonian to configurations
of discrete variables, one finds linear dependencies between
the counts of certain local patterns and and certain other ones.  
(Here ``local pattern'' means a particular combination 
of two or more discrete variables in nearby sites.)
Thus, if every local pattern is allotted an independent 
coupling coefficient, the values of these coefficients
will be ill-determined: a whole family of parameter sets
gives identical energies for all possible configurations.
Such dependencies arise when the discrete variables are
Ising spins~\cite{uzi} or tiles (``decorated'' by atoms)
in quasicrystal-related phases~\cite{MgZn}; they arose
in the previous fit of a cluster 
Hamiltonian~\cite{brommer2007} and were handled by
arbitarily setting certain coefficients to zero.

In fact, interacting clusters are the simplest
of the cases where these dependencies are observed.
Our task is to write
   \begin{equation}
       E_{\alpha\beta} = E_0 + U_\alpha +  U'_\beta + V_{\alpha\beta}.
   \label{eq:Epair-breakup}
   \end{equation}
The difficulty is that the solutions of \eqref{eq:Epair-breakup}
are ill-defined, since it is invariant under
   \begin{subequations}
  \label{eq:indetermined}
  \begin{eqnarray}
  \label{eq:indetermined-V}
      V_{\alpha\beta} &\to& 
   V_{\alpha\beta} - A_\alpha - A'_\beta; \\
  \label{eq:indetermined-U-L}
     U_\alpha &\to& U_\alpha+  A_\alpha; \\
  \label{eq:indetermined-U-R}
     U'_\beta &\to& U'_\beta+  A'_\beta,
  \end{eqnarray}
   \end{subequations}
where $\{ A_\alpha\}$ is an arbitrary set of energy shifts.
Within an extended crystal, with bonds oriented 
in several directions, there will be different functions
$A_\alpha$ and $A'_\alpha$ associated with each direction of bond;
the one-body term gets shifted by the sum of all these.
Thus, even if we constrained all $U_\alpha\equiv 0$, that
would not resolve the indeterminacy in $\{ V_{\alpha\beta} \}$
There is a similar indeterminacy in the single-cluster term:
   \begin{subequations}
  \label{eq:indetermined-U-single}
  \begin{eqnarray}
   U_\alpha &\to& U_\alpha - B; \\
   U'_\beta&\to& U'_\beta- B'; \\
   E_0  &\to& E_0 + B+B'.
  \end{eqnarray}
   \end{subequations}

We do resolve it by imposing the simple rule 
   \begin{subequations}
   \begin{eqnarray}
    \label{eq:fix-gauge-V}
    \sum_\alpha V_{\alpha\beta}
  = \sum_\beta V_{\alpha\beta} &\equiv& 0;\\
    \label{eq:fix-gauge-U}
   \sum_\alpha U_\alpha \equiv \sum _\beta U'_\beta &\equiv& 0.
   \end{eqnarray}
   \end{subequations}
Given an initial set $E_{\alpha\beta}$, it is
easy to implement Eq.~\eqref{eq:fix-gauge-V}
using Eqs.~\eqref{eq:indetermined} with
   \begin{subequations}
   \begin{eqnarray}
       A_\alpha &\equiv& \frac{1}{m}(\sum_\beta E_{\alpha\beta}); \\
       A'_\beta &\equiv& \frac{1}{m}(\sum_\alpha E_{\alpha\beta}).
   \end{eqnarray}
   \end{subequations}
After that we enforce Eq.~\eqref{eq:fix-gauge-U}
by applying \eqref{eq:indetermined-U-single}
with $B\equiv (\sum_\alpha U_\alpha)/m$ and similarly $B'$.

\subsection{Singular Value Decomposition methods}
\label{sec:SVD-method}

The matrix of interactions $V_{\alpha\beta}$ for the
discrete orientations has $m^2$ entries and
this implies a large number of fit parameters, even after 
symmetries are taken into account and redundancies eliminated.  
For example, Ref.~\onlinecite{brommer2007}  --
taking the discrete approach --
fitted a separate interaction parameter 
for every inequivalent combination of orientations 
of the nearest or second-nearest cluster neighbors,
which amounted to 42 interactions (slightly fewer
after removing redundancies).

However, we anticipate that some combinations
of these parameters are unimportant.
Imagine, e.g., if a cluster developed a charge $q_\alpha$
dependent on its orientation, then the cluster orientation
would be a Coulomb interaction proportional to
$q_\alpha q_\beta/|R_{ij}|$ and thus of completely separable
form.  To parametrize such an interaction for $m$ orientations, 
we do not need $O(m^2)$ numbers, but only one overall strength 
and a list of $O(m)$ charge strengths $q_\alpha$.
An electric dipole interaction is similar, but since the
dipole moment has three components, the interaction
would take the form

Our recipe to reduce parameters is to posit that the
interaction matrices can be put in the form
   \begin{equation}
    V_{\alpha\beta} = \sum _{\mu=1}^{m'} \sigma_\mu g^\mu_\alpha g'^\mu_\beta.
   \label{eq:SVD-form}
   \end{equation}
The implicit idea here is that the clusters are weakly perturbing
some ``field'' which fills the space between them.
Think of $g^\mu_\alpha$ as the ``charge'' presented by the first cluster
when it is in orientation $\alpha$, similarly $g'^\mu_\alpha$ as the
``charge'' presented by the second cluster, while $\sigma_\mu$ is
the effective interaction of these charges (expected to decay with
separation).  More exactly, $\mu$ indexes different kinds of ``charge'',
or different components of the same charge.  For example, if
the tetrahedron atoms carried a literal charge, then $\mu$ would
index the cluster's moments (monopole, dipole, etc) and their
tensor components (three for the dipole case).
In CaCd$_6$, we expect the interaction is
mediated by elastic strains of the intervening atoms,
or possibly by direct couplings (which are represented by
the oscillating tail in the pair potentials shown in
Figure~\ref{fig:EOPP-CaCd}, or analogous tails
in the EAM potentials~\cite{brommer2006}).

\SAVE{In general, the dipole interactions would make
a $3\times 3$ matrix; assume here that one of the Cartesian axes
includes the inter-cluster vector ${\bf R}_{ij}$ -- then  the dipole
interaction has no cross-terms between different components.}

\SAVE{Similarly, if the interaction were mediated
only through a modulation of the conduction electron density;
in a metal such modulations oscillate at wavevector $2k_F$ 
(where $k_F$ is the Fermi wavevector), i.e. the entire
information transmitted to a distant second cluster is 
incorporated in two components of a complex amplitude 
(equivalently a single amplitude and a phase angle). 
The interaction of each cluster with this modulation parameter
is a complex coupling $g_\alpha$ so the effective cluster
interaction must take the form 
$V_{\alpha\beta} \propto {\rm Re} [(g_\alpha^*) g'_\beta]$.}

For definiteness, let's augment \eqref{eq:SVD-form},
by an  orthonormal condition $\sum _\alpha g_{\mu\alpha}
g_{\nu\beta} = \delta_{\mu\nu}$.  Then, these two equations
are equivalent to the {\it singular value decomposition}, 
or in matrix notation
   \begin{equation}
       \ulV = \ulg^T\; \ulsigma\;  \ulg'
   \end{equation}
where $\ulsigma$ is the diagonal matrix 
$\ulg$ and $\ulg'$ are orthogonal matrices of singular values
(which we take to be nonnegative and decreasing).
In this form, we have $m'=m$~\footnote{
Actually, our resolution of  the redundancy is equivalent to
projecting away the subspace in $V_{\alpha\beta}$ parallel
to $(1,1,1,1...)$ so its rank was reduced by one. 
Therefore, the last singular value $\sigma_m$ is always zero.
}
But our expectation (to be checked!) is that the singular
values $\sigma_\mu$ rapidly get smaller, so we can truncate
\eqref{eq:SVD-form} after (say) three terms and reduce
the 42 parameters to a handful.

In practice this decomposition is much more powerful when
symmetries come into play.  Following the analogies to
(electric or elastic) dipoles and quadrupoles,
we imagine that the dependence of the ``charge''
on orientation $\Omega$ could be written as a
smooth function $g(\Omega)$ with the angular dependence
of an orientational harmonic.  This idea is tested
in Section~\ref{sec:results-continuous}

\section{Results: discrete orientations}
\label{sec:results-discrete}

In this section, we use the framework of Section~\ref{sec:framework} 
to fit a cluster potential for the twelve discrete, 
symmetry-related optimal orientations, similar to the form used
in Ref.~\onlinecite{brommer2007}.  
Our microscopic Hamiltonian is defined by EAM potentials of Ref.~\onlinecite{brommer2006}.
For each relaxation, we use an initial structure in which 
the ``regular'' atoms (all atoms
excluding the tetrahedra) are in their averaged positions
with cubic crystal symmetry (space group $Im\bar{3}$), 
as refined in Ref.~\onlinecite{gomez2003} 
(this structure has
the Pearson symbol cI208).
Each tetrahedron is initially regular 
(having radius 1.775\AA~ and centered on the 
Tsai cluster center) and in the prescribed orientation $\ulO_i$.

\SAVE{The dodecahedron sites have large variations as the tetrahedra are
rotated, which should show up as a large DW factor for these atoms
in the Gomez-Lidin refinement.}

\subsection{Cluster pair interaction}
\label{sec:cluster-pair-interaction}

\SAVE{
For the zero strain data in this section, and also the 
initial data for the pressure runs,
we probed all 12 backgrounds ignoring symmetries.
In the pressurized relaxations, for faster computation, symmetries were
taken into account, at least partly.
We did an  independent analysis for each possible background $\ulO_{bg}$.
For $b$-bonds, identical configurations after mirror operations 
$m_x$, $m_y$, and $m_z$, and combinations of them are used. 
For $c$-bonds, the threefold rotation and inversion are used.}

\SAVE{For discrete fits, the presented potentials
are averaged over all backgrounds.  For the continuous fits,
though, unaveraged potentials are shown.}

\begin{table}
  \centering
  \begin{tabular}{c |  c  c  c  c |}
       &$\dir{X}{+}{r}$ & $\dir{X}{-}{r}$& $\dir{X}{+}{l}$& $\dir{X}{-}{l}$\\
\hline
$\dir{X}{+}{r}$ &  $-2.8$&  0.5&  4.2&  $-0.4$\\
$\dir{Y}{+}{r}$&  $-8.6$&  5.7&  $-1.9$&  3.9\\
$\dir{Z}{+}{r}$&  $-5.2$&  2.0&  $-1.1$&  3.9\\
$\dir{X}{-}{r}$&  $-6.8$&  3.9&  $-7.3$&  4.2\\
$\dir{Y}{-}{r}$&    6.0&  $-6.3$&  3.2&  $-1.1$\\
$\dir{Z}{-}{r}$&    0.3&   2.6&    3.2&  $-1.9$\\
$\dir{X}{+}{l}$&  $-5.3$&   3.5   & 3.9  &   0.5   \\
$\dir{Y}{+}{l}$& $-2.3$ &   11.7  & 2.6  &   2.0   \\  
$\dir{Z}{+}{l}$&  $-17.6$&   11.7  & $-6.3$ &    5.7  \\
$\dir{X}{-}{l}$&   9.6  &  $-5.3$ & $-6.8$ &   $-2.8$  \\
$\dir{Y}{-}{l}$&  21.3  &   $-17.6$ &  0.3 &   $-5.2$  \\
$\dir{Z}{-}{l}$&  21.3  &   $-12.3$ &  6.0 &   $-8.6$  \\
\hline
  \end{tabular}
  \caption{Interaction matrix for two clusters separated
by $[1/2,1/2,1/2]$ (=$c$-linkage), using EAM potentials (in meV).
Due to the threefold symmetry of this linkange,
all interactions are invariant under cyclic permutations
$(xyz)$ in the labels when applied to both clusters; thus columns 
are given here only for the ``$X$'' orientations.
\SAVE{CLH has implemented the conversion from Woosong convention
to official one, by switching r/l whenever $-$.}
}
  \label{tab:cbondInteraction}
\end{table}

We computed interactions for two kinds of cluster pairs,
separated by the vector [1/2,1/2,1/2] (=$c$ linkage)
or [0.0,1] (=$b$ linkage), placed in a \supcell{3}{3}{3}
supercell.  We used the uniform background method and averaged
over the twelve possible backgrounds; as requiared by 
symmetry, the averaged single-body term was zero
(i.e. all orientations were equivalent). 
The resulting pair interactions are given in
Tables \ref{tab:cbondInteraction} and
\ref{tab:bbondInteraction}.

To make sense of the computed interactions, it is 
essential to consider the local symmetry of the
cluster pair, which is reflected in the symmetry
of $V_{\alpha\beta}$ as laid out in Sec.~\ref{sec:interaction-symmetries}.

For the $c$ pair,  the point group (centered on the linkage's midpoint)
$\bar{3}$ ($D_3$); the subgroup which maps each cluster to itself is
$3$ ($C_3$).  This threefold rotation, in terms of our orientation labels, 
just cyclically permutes $(XYZ)$ without changing the $\pm$ or $r/l$ part
of the labels.  As for the $b$ pair of clusters, 
its point group is $2/mm$ ($D_{2h}$) and the subgroup that 
maps each cluster to itself is $2m$ ($C_{2v}$) which has order 4.
Its generators are the mirror operations $m_x$ and $m_y$, which 
respectively invert the $x$ or the $y$ coordinates.  (Remember
we singled out the $z$ direction by aligning the pair with it.)
The action of $m_x$, for example, is to switch $+X \leftrightarrow -X$,
not switch the sign of the $Y$ or $Z$ orientations, and in all cases
to switch $r \leftrightarrow l$.
Finally, in all cases it is convenient to let inversion be the
fundamental operation that swaps the two clusters; its action on
our labels is to always switch $+ \leftarrow -$ and always
switch $r \leftrightarrow l$.  (Note that any proper rotation
always preserves the $r/l$ indices whereas any improper rotation
always switches them.)  We take advantage of these symmetries 
to reduce the number of columns displayed in the interaction 
matrices.

\LATER{Which cluster is which, in the interaction matrix?}

\subsubsection{Justification of functional form}
\label{sec:justify-form}

\LATER{Combine these observations.
magnitudes of the terms fall into groups which depend on
the main orientations, and not the $\pm$ or $r/l$ attribute of the orientation.
In the (001) linkage interaction, the 
$X$--$X$ terms are large, about $\pm 0.017$ eV,
the $X$--$Y$ terms are second largest, about $\pm 0.007$ eV,
the $Y$--$Y$ and $Z$--$Z$ terms are about $\pm 0.003$ and $0.004$ eV,
and all other terms are much smaller.
The significant interactions change sign under $+X/-X$, $+Z/-Z$, and $Y_r/Y_l$
ut not under $+Y/-Y$, $X_r/X_l$, or $Y_r/Y_l$.  
Is it valid to say the $X/Y/Z$, $\pm$, and $r/l$
labels have influence on the energy,  in that order?
In any case, explain the hierarchy by considering how the actual positions of
tetrahedron atoms look in relation to the (001) direction? }

\LATER{Looking at Tables \ref{tab:bbondInteraction} 
and \ref{tab:cbondInteraction}, we can justify the relative
values of the interactions for some orientations.
For example, suppose that we have a $\dir{X}{+}{r}$ tetrahedron at $(0,0,0)$ 
then, the tetrahedron at $(0,0,1)$ would highly prefer to be $\dir{X}{-}{r}$ or $\dir{X}{-}{l}$. }

Can we make sense of the patterns of interactions?
First of all, we see that just a few combinations have much larger interactions
than any others.  In the $c$-linkage case, for example, the
pair $(\dir{Z}{-}{l},\dir{X}{+}{r})$ and permutations
has interaction 21.3 meV; due to inversion symmetry relating
the two clusters [we are using \eqref{eq:V-swap}],
this has the same interaction as the pair $(\dir{X}{-}{l},\dir{Z}{+}{r})$.
In turn that is equivalent by cyclic permutation to
$(\dir{Y}{-}{l},\dir{X}{+}{r})$  in our table, which is indeed 21.3 meV.
The interaction $(\dir{Y}{-}{l},\dir{X}{-}{r})$ is almost as big, 
$-17.6$; this is equivalent to 
The interaction $(\dir{X}{+}{l},\dir{Y}{+}{r})$, i.e. 
$(\dir{Z}{+}{l},\dir{X}{+}{r})$ in the table, by inversion.
Overall, it can be seen that the $r$ orientations have
stronger interactions than the $l$ orientations.

\begin{table}
  \centering
  \begin{tabular}{c |  c  c  c  c  c  c| }
&$\dir{X}{+}{r}$ & $\dir{X}{-}{l}$ & $\dir{Y}{+}{r}$ & $\dir{Y}{-}{l}$ & 
                                     $\dir{Z}{+}{r}$ & $\dir{Z}{-}{l}$ \\
\hline
$\dir{X}{+}{r}$ &  17.4 & $-17.0$ & 6.7 & 6.5 & $-1.1$ & 0.8\\
$\dir{X}{+}{l}$ &  18.7 & $-16.5$ & 6.5 & 6.7 & $-0.6$ & 0.4\\
$\dir{X}{-}{l}$ &  $-17.0$ & 17.4 & $-7.0$ & $-8.5$ & $-0.6$ & 0.4\\
$\dir{X}{-}{r}$ &  $-16.5$ & 18.7 & $-8.5$ & $-7.0$ & $-1.1$ & 0.8\\
$\dir{Y}{+}{r}$ & $-8.5$ & 6.5 & $-2.5$ & $-2.3$ & 0.6 & $-0.7$\\
$\dir{Y}{+}{l}$ &  6.5 & $-8.5$ & 4.4 & 2.8 & 1.4 & $-1.4$\\
$\dir{Y}{-}{r}$ & 7.0    &  6.7   &  $-2.3$ &  $-2.5$ &  1.4   &  $-1.4$ \\
$\dir{Y}{-}{r}$ & 6.7    & $-7.0$ &    2.8  &   4.4   &  0.6   &  $-0.7$ \\
$\dir{Z}{+}{r}$ &  0.4    & 0.8    &  $-1.4$ &  $-0.7$ & $-3.8$ &   4.1   \\
$\dir{Z}{+}{l}$ &  0.     & 0.4    &  $-0.7$ &  $-1.4$ & $-2.6$ &   4.0   \\          
$\dir{Z}{-}{l}$ & $-0.6$  & $-1.1$ &    1.4  &  0.6    &  2.6   &  $-3.8$ \\          
$\dir{Z}{-}{r}$ & $-1.1$  & $-0.6$ &    0.6  &  1.4    &  3.0   &  $-2.6$ \\
\hline
  \end{tabular}
  \caption{Interaction matrix $V_{\alpha\beta}$
for two clusters related by $[0,0,1]$  vector (=$b$-linkage), in meV.
In all interaction matrices the cluster 1 orientation
is given in column headings, and cluster 2 orientation by
row headings.  
\TODO{WC: is the preceding correct, or is it the other way around?}
The interaction is invariant if one switches
r$\leftrightarrow$l in {\it both} labels, so the columns with ``$+l$'' 
and ``$-r$'' labels are omitted.
\SAVE{I have switced r/l in labels with $-$ so as to convert from
Woosong's scheme to the official one.}
\LATER{CLH:  convert to $d$ format? but the $d$ format seems to put 
``\$...\$'' around the entry AND requires a decimal point.}
}
  \label{tab:bbondInteraction}
\end{table}

Similarly, in the $b$ interactions (Table~\ref{tab:bbondInteraction}),
we that $X$ orientations 
(of all flavors) have by far the largest interactions, while $Z$
orientations have the least.

\TODO{Woosong: An important property I expected from thes matrices is that
the rows and columns add up to zero.  That doesn't seem to be the case!
What's going on here?}

If the clusters were far apart and weakly perturbing their surroundings,
we would expect the inversion of a cluster would have exactly the
opposite coupling, in the case of a tetrahedron, since it moves
the atoms to the places complementary to them.  To say this another way,
consider the union of a tetrahedron and its inverse, i.e. a cube:
whatever is the lowest nonzero ``multipole moment'' of the tetrahedron,
we would expect that one to be zero in the case of the cube.
Thus, we expect that {\it approximately} $V_{\alpha\beta}=-V_{\alpha'\beta}$
when $\alpha$ and $\beta$ are related by inversion.
This expectation is borne out in the $b$-bond interactions, as can be
seen by comparing adjacent entries in columns 1 and 2, 3 and 4, or 5 and 6.
However, in the case of the $c$-linkage it does not work well:
the largest interactions 
(($-17.6$ and $+21.3$) are related (see the leftmost column) by
$\dir{Z}{+}{l}\leftrightarrow\dir{Z}{-}{l}$.

\LATER{Talk about idea of an isotropic elastic medium.
Then the interaction would be same under a 90 degree rotation about z
-- which is actually possible, but not a symmetry.  
No, that doesn't work -- proving that the particular arrangement
of the cage matters.}

A somewhat surprising fact is that
the size of $c$-bond interactions is
the same as that of the $b$-bond interactions;
this is confirmed (see below) by the fact the
respective dominant singular values from each interaction 
are practically the same.
The best absolute measures, the matrix norms, 
are respectively 94.7 and 83.1, for a ratio $\sim 1.1$.
One would have expected the $c$-bond interaction to be
larger -- by a factor of $\sim 1.5$,
if the interaction simply scaled as $1/R^3$.

\subsubsection{Singular Value Decomposition}
\label{sec:results-SVD}

\SAVE{ For the $b$-bond, I believe the
point group of the pair is $2/mm$ ($D_{2h}$) or for one cluster, $2m$ ($C_{2v}$).
The irreps are named $A_1$ (trivial), $A_2$, $B_1$, $B_2$.
For the $c$-bond, I believe the point group of the pair 
is $\bar{3}$ ($D_3$) or for one cluster, $3$ ($C_3$).
The irreps are named $A$ (trivial) and $E$.}

\SAVE{Useful online sources for the group theory.
(1) Wikipedia, ``List of character tables for chemically important 3D point groups''.
(2) UC Davis ChemWiki, ``Group Theory: Theory''
(3) UCL lectures by Andrew Wills on group theory.}

\begin{table}
  \centering
  \begin{tabular}{c | c |   c c c c }
 $\sigma$  (meV) & irrep & 
    $\dir{X}{+}{r}$& $\dir{X}{-}{l}$& $\dir{X}{+}{l}$& $\dir{X}{-}{r}$\\
\hline
$81.51$            & A & $-0.1380$ & $-0.0054$ & $-0.3181$ & $0.4616$ \\
$33.79$            & E & $0.2339$ & $0.4564$ & $0.4140$ & $0.4819$ \\
                   &   &  $-11.7^\circ$  & $164.9^\circ$ & $-66.9^\circ$ & $138.0^\circ$ \\
$5.443$            & A & $0.4204$ & $0.0739$ & $-0.3675$ & $-0.1268$ \\
$2.278$            & E & $0.7060$ & $0.1414$ & $0.2760$ & $0.2684$ \\
                   &   & $177.1^\circ$ & $-68.0^\circ$ & $-109.8^\circ$ & $161.6^\circ$ \\
$1.385$            & E & $0.2987$ & $0.2812$ & $0.5896$ & $0.3883$  \\
                   &   & $142.7^\circ$ & $176.5^\circ$ & $-68.5^\circ$ & $-79.5^\circ$ \\
$1.330$            & A & $-0.2329$ & $0.4944$ & $-0.1170$ & $-0.1445$ \\
$3.288\times 10^{-2}$&E & $0.1557$ & $0.5994$ & $0.2673$ & $0.4600$ \\
                   &  &  $33.4^\circ$ & $65.8^\circ$ & $55.8^\circ$ & $-56.0^\circ$ \\
\hline
\end{tabular}
\caption{
\SAVE{Irreps sent by WC on 8/15/12.}
Singular values and right singular vectors for $c$-linkage.
This cluster pair has a symmetry under cyclic permutations 
$(XYZ)$ and consequently every singular vector falls into
one of the two representations of that symmetry group.
Those labeled I follow the identity representation; the omitted
columns for $Y$ and $Z$ type orientations have the same amplitudes
as the corresponding (shown) columns for $X$ type orientations.
The representations labeled 2 are twofold degenerate, and take the
form $A \cos \phi_X$, $A\cos(\phi_X+120^\circ)$, and 
$A\cos(\phi_X+240^\circ)$ for the $X$, $Y$, and $Z$ entries,
where $(A,\phi_X)$ is shown in the table;  a second, degenerate
singular vector takes the same form but with $90^\circ$ added
to all the phases.
\TODO{Shift the phase angles for the complex (dimension 2) representation
w.l.o.g. so as to start at $\phi_X=0$.}
}
\label{tab:cbondSVD}
\end{table}

Applying the singular value decomposition analysis to 
the data in Tables \ref{tab:bbondInteraction} and \ref{tab:cbondInteraction}.
yields the results in Tables \ref{tab:bbondSVD} and \ref{tab:cbondSVD}.
Again, these need to be understood in light of the point symmetries.
In particular, every singular vector must transform under one of
the irreducible group representations, which 
have been identified and indicated in the tables.
The irreps of $3$ are named $A$ (trivial) and $E$ (two-dimensional);
the irreps of $2/m$ are named $A_1$ (trivial), $A_2$, $B_1$, and $B_2$.

Having identified the largest entries in the interaction matrix,
we can interpret the leading singular values and their singular vectors.
For example, let's approximate the $b$-linkage interaction matrix
by keeping only its largest terms, found in the $4\times 4$ subblock 
represented by the upper left eight entries in Table~\ref{tab:bbondInteraction}).
They are all close to $\pm 17$meV, with a pattern of signs that
gives a rank-1 matrix.  There would be just one nonzero singular value
$\sigma= 4(17)$meV = 68 meV. Its singular vector would have four
entries $\pm 1/2$, for the four kinds of $X$ orientation, and zero
on all others. The pattern of signs is $(+-+-)$ in the convention of
Table~\ref{tab:bbondSVD}, which corresponds to the $B_2$ representation;
in this approximation the top row of the table would read 
$(-0.5, 0,0,0)$ and we see this is a good approximation of the
actual singular vector.  Similarly, the second singular vector
in Table~\ref{tab:bbondSVD}
is approximately ($0,0,0.5, -0.5)$, expressing the difference
between $Z+$ and $-Z$ orientations.
\LATER{A given $X$ or $Y$ tetrahedron has atoms with 
nonzero components $\pm z$; although exactly opposite,
the interactions do not exactly cancel because one of them
is closer to the other cluster.
I guess the admixture of the $X$ and
$Y$ orientations in the actual vector is depedent on
the rotation represented by the $l$ and $r$ labels.}

It can also noted that, for the $b$ interaction,
 both of the two leading singular vectors treat the $r$ and $l$ variants 
orientations the same; if we retained only these two terms,
that truncated interaction sees only only six kinds of orientation, 
with the $r$ and $l$ variants lumped together.
On the other hand, the leading singular vector for the $c$ interaction
is very far from treating $r$ and $l$ equivalently.

\LATER{Our analysis is somewhat incomplete, in that we do not 
check the relationship between the right and left singular vectors,
required by the inversion operation that swaps them.  
A given term in \eqref{eq:SVD-form} must be invariant 
under the swap \eqref{eq:V-swap}.  Since the standard
singular-value decomposition always gives positive singular 
values, the consequence is that the right singular vector 
(after the permutation induced by inversion) is either
the same as the left one, or its negative.  We would
prefer to take the convention it is the same, and allow
$\sigma_\alpha$ to have either sign.}

\begin{table}
  \centering
  \begin{tabular}{c | c | c c c c }
\hline
$\sigma$ (meV)  & irrep & $\dir{X}{+}{r}$&$\dir{Y}{+}{r}$&$\dir{Z}{+}{r}$&$\dir{Z}{-}{l}$\\
    \hline
$81.45$ & $A_2$ & 0.4621&  $-0.1908$&  $0$ & $0$\\
$15.59$ & $A_1$ &  $0.1282$& $-0.1625$& $0.4881$& $-0.4196$\\
$3.836$ & $B_1$ & $0.2402$&$0.3507$&$0.2705$& $-0.2557$\\
$3.212$ & $B_2$ & 0.3727&0.3333&  0& 0 \\
$2.745$ & $A_1$ & 0.3780&$-0.2321$&$-0.3247$&0.0329\\
$1.375$ & $B_1$ & $0.0484$ &0.0077&0.4245&0.5612\\
$2.733\times 10^{-1}$ & $A_1$ & $-0.0858$&$-0.2940$&0.2701&0.4894\\
$2.073\times 10^{-1}$ & $B_1$ &  0.3552&$-0.3313$&0.1046&$-0.1312$ \\
$1.364\times 10^{-1}$ & $B_2$ & $-0.3333$ & $0.3727$&  0&  0\\
$1.307\times 10^{-1}$ & $A_2$ & $0.1908$ & $0.4622$& 0& 0\\
$2.017\times 10^{-2}$ & $B_1$ & $-0.2526$&$-0.1308$&0.4854& $-0.3200$\\
\hline
 & irrep & signs $X$ & signs $Y$  & signs $\dir{Z}{+}{r/l}$ & signs $\dir{Z}{-}{l/r}$ \\
\hline
&  $A_1$  &  $++++$& $++++$&  $++$& $++$ \\
&  $B_1$  &  $+--+$&  $+--+$&  $+-$& $-+$ \\
&  $A_2$  &  $++--$&  $+-+-$&  $00$& $00$ \\
&  $B_2$ &  $+-+-$&  $++--$&  $00$& $00$  \\
\hline
  \end{tabular}
  \caption{Singular values $\sigma$ and right singular vectors for $b$-linkage.
The singular vectors belong to representations of the 
point group $2m$ (=$C_{2v}$), as given in the second column: 
\SOON{Need to correct sign of last $z$ column to refer to $r$ entry,
it was the $l$ entry (what Woosong called $r$)}
\SAVE{Old notation I, A, B,C = $A_1$ (trivial), $A_2$, $B_1$, $B_2$.}
Each singular vector has twelve entries, one for each 
of the orientations (column labels).
\SAVE{$A_2$ is the one which is $+1$ for proper and $-1$ for improper,
ergo $+--+$}
The four orientations beginning with direction ``$X$'' or ``$Y$'', 
e.g. ``($\dir{X}{+}{r},\dir{X}{+}{l},\dir{X}{-}{l},\dir{X}{-}{r})$'', 
always have the same amplitudes
apart from sign, so only the first column is printed from each group of four.
Similarly, in each of the pairs ``($\dir{Z}{+}{r},\dir{Z}{-}{r})$'' and the
and `$\dir{Z}{+}{l},\dir{Z}{-}{l})$'' only one of the two is needed.
The pattern of relative signs for the group of four or the pair 
depends on the group representation,
as given at the bottom of the table.
\SAVE{12/18/12 CLH swapped $r/l$ for the ``$-$'' entries 
to convert from Woosong's convention.}
}
  \label{tab:bbondSVD}
\end{table}

\subsection{Testing validity of discrete effective Hamiltonian}
\label{sec:test-validity-discrete}

\subsubsection{Checks from the fitting process}
\label{sec:check-fit-residuals}

We have checked the validity of the orientation potentials in
multiple ways.  First, we report the results of two checks which 
are inherent to the uniform-background scheme for fitting, as 
explained in Sec.~\ref{sec:uniform-background}.

The first check is that, in the fit with 
a particular backgroun, the apparent one-body energy vector entirely 
represents interactions with the background orientation, and therefore
has a low symmetry.  
\SAVE{As noted in Sec.~\ref{sec:uniform-background}, the physical one-cluster 
terms $\UC(\ulO)$ are the same for any one of the twelve optimal orientations, 
so they contribute only to the {\it constant} term in the
discrete orientation Hamiltonian we are extracting.} 
The apparent one-body energy varies between $+0.03$eV and $-0.05$eV for most backgrounds.
\SAVE{Similar offsets, with the same energy scale,
may be seen in Figure~\ref{fig:ContinuousFit} where
certain angles that ought to be symmetry equivalent, give different single-body
energies on account of the uniform background.}

Next, from the fitted pair interactions of our chosen clusters, we can predict
the summed pair interactions of either cluster with its 13 $[100]$
and $[1/2,1/2,1/2]$ neighbors in the fixed background, for each of the
fixed backgrounds.  The agreement is excellent: the root-mean-square
residual is 2.6 meV when the varied clusters are a 
$[1/2,1/2,1/2]$ pair,
or 2.3 eV when they are a $[100]$ pair.
This residual is
due to either multi-body interactions or to more distant neighbor
interactions (with the background clusters).
\SAVE{WC sent the fitted one-body interaction for each
background, in the form of plots showing the fitted "single body"
energy as well as the predicted "single body" energy, as a function
of the 12 orientations.  
See folder "prog/qx/woosong/Results/E-pair/Singlebody_estimates-figs".
In those plots, red lines from the interaction matrix rows [left panel]
columns [right panel]; blue dotted lines from the two-body interaction.}

Incidentally, in a preliminary study, we directly tested
the assumption that farther neighbor interactions are
significantly weaker, by directly fitting $V_{\alpha\beta}$
trial fits of the pair interaction
and therefore could be omitted. 
for second nearest-neighbors displaced by the vectors $[0,1,1]$ and also
$\frac{1}{2}[1,1,3]$.
(This was only carried out using a neutral background of Ca-filled clusters, 
as mentioned in section ~\ref{sec:uniform-background}.)
These interaction matrices gave singular values more than an order of magnitude smaller 
than those of the nearest-neighbor interactions.

\TODO{CLH:  WC sent info on the further neighbor in email 8/16/12, see woosong-Cd4.in0815.
The old results mentioned were with Ca filled clusters.}

The second check from fitting is that,
insofar as multi-cluster interactions can be neglected,
different backgrounds should give an {\it exactly} identical interaction
matrix --  even if there are longer range pair interactions,
We found that this is largely so.
To quantify the dependence on background
we considered how a particular element $V_{\alpha\beta}$
of the $12\times 12$ interaction matrix varies while
we cycle $\ulO_{bg}$ through all 12 possible  values,
and compute the standard deviation  of $V_{\alpha\beta}$.

In the case of the [1/2,1/2,1/2] cluster linkage,
the average standard deviation is about
1--2 meV.  The larger interactions 
and larger standard deviations were between
any tetrahedron of orientation $\dir{X/Y/Z}{+}{r}$ and any one of
orientation $\dir{X/Y/Z}{-}{l}$.

In the case of the [1,0,0] linkage, the standard
deviation was typically $\sim 0.6$ meV), i.e.
half as big as for the $c$-linkage.
deviation was larger than average for the $4^2$ interaction terms that
combine $X$--$X$ type orientations 
by a factor of almost two, and smaller by a factor of nearly two in
the $8^2$ terms not involving any $X$ type orientation.
(The $X$--$X$ pairs had by far the strongest interactions, too),

To summarize, the multi-cluster interactions, as measured
by the standard deviation of $V_{\alpha\beta}$ while the
background is varied, is $\sim$15\% of the 
cluster pair interaction, for both kinds of linkage.
Thus, the multi-cluster interactions are significant,
but an order of magnitude down from the dominant pair interactions.

\subsubsection{Tests by direct comparison with total energies}

\CLH{This test, as well as the MC simulations,
were done using the full pair interaction matrix, not
a simplified version derived using SVD.}

\TODO{Woosong: for later, I would like to get the data sets
plotted in the scatter plots, for the purpose of extracting and
reporting the standard deviation from the fit line.}

\TODO{Woosong: I just noticed the x axes of the scatter plots.
Where it says ``EAM Energy (eV) $- 5.091\times 10^3$'' shall
I infer you subtracted that constant term from the total?}

To fully validate the effective Hamiltonian fitted from
uniform backgrounds (in \supcell{3}{3}{3} supercells),
we should test it on databases with 
and the energies computed with 
configurations other than the fitting database.
Our test datasets was a \supcell{3}{3}{3} supercell with
in which randomly oriented tetrahedra were placed, and then relaxed
in two ways (see Sec.~\ref{sec:constrained-relax}), giving two variant databsets:
\begin{itemize}
\item[(1)]
relaxed with the continuous orientation $\ulO_i$ held strictly fixed;
\item[(2)]
the same random orientations, unconstrained relaxation  so the orientation can change
\end{itemize}
(In the latter case, it was assumed without checking that
every cluster's orientation remains in the discrete well belonging to 
optimal orientation with which it was initialized.)

\begin{figure}[h!]
  \begin{center}
    \includegraphics[width=0.8\columnwidth]{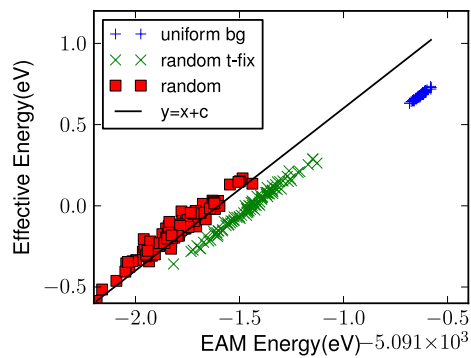}
  \end{center}
  \caption{Scatter plot, showing effective Hamiltonian
prediction compared to the relaxed energies,
for a $3\times3\times 3$ superlattice
with each cluster in a random discrete orientation (out of the
twelve optimal ones).
Different symbols refer to different conditions of the fit:
$+$ = the uniform background configurations 
(constrained relaxation) used for the fit;
$\times$ = random configurations relaxed with the 
constraints (Sec.~\ref{sec:constrained-relax})
and $\square$ = relaxed unconstrained.
These last are offset to the left,
due to the energy reduction
when tetrahedra (under forces from neighboring tetrahedra)
relax to orientations slightly different from the discrete
minima of the single-tetrahedron orientational potential.
\SAVE{The constant term in the fit is the reason
that the line in this figure doesn't pass through the origin.}
\SAVE{In older versions of the figure:
x-bbond = pairs in (1,0,0) neighbors; bc-bonds are in 
(1,1,3)/2 neighbors; ``test config'' WC did not remember.}
}
\label{fig:EffectiveHamiltonian}
\end{figure}

Figure \ref{fig:EffectiveHamiltonian} shows scatter plots
comparing the effective Hamiltonian prediction (vertical axis)
with the actual EAM interaction, for the two random datasets
and also the input fitting dataset.  An offset of about 
$+94.30$ eV/cluster has been subtracted, representing the
EAM energy of all the atoms in one primitive cell.
\footnote{Interestingly, Figure \ref{fig:EffectiveHamiltonian}
demonstrates that any uniform background state is about 
20 meV/cluster higher in energy than a typical random pattern.}
A systematic constant offset is visible due to the use 
of constrained relaxation.  It should be noted that 
this is entirely canceled out in way we fit the pair 
interactions (see Sec.~\ref{sec:redundancies}.

\LATER{Less obviously, out of the constrained points,
the random-orientation kind ($\times$) have a slightly
lower pair potential energy than the uniform-background
kind.  CLH speculates, in an irregular
background there are more relaxations possible  that
distort (but do not rotate) the clusters and thereby
lower energy.}
\SAVE{WC speculates that  the uniform-background points fall
on a different line than the constrained-random ones
because the three (or more) body interaction terms tend to
add up, whereas they tend to cancel in the random case.
CLH is not so sure.}

\begin{figure}[h!]
  \begin{center}
    \includegraphics[width=0.8\columnwidth]{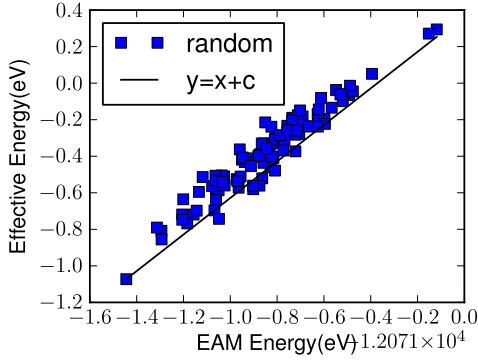}
  \end{center}
  \caption{Scatter plot, comparing energies of relaxed  
atomic configurations in $4\times 4\times 4$ supercells
to those predicted by the discrete cluster Hamiltonian, 
which was fitted from a smaller ($3\times 3 \times 3$) supercell.
}
  \label{fig:EffectiveHamiltonian4x4x4}
\end{figure}

The scatter of the total energy appears to be a bit under $\pm 0.1 $eV,
coming from $2\times 3^3=54$ clusters, or a total
of $54\times 14/2= 378$ linkages, since each has $(8+6)$
$c$- and $b$-type neighbors.  If we imagine that each interaction
deviates deviates randomly and independently from the fitted
energy, this random energy comes  to $\sim 4$ meV/linkage,
consistent in magnitude with the residuals estimated just
above (in Sec.~\ref{sec:check-fit-residuals}).
\SAVE{The scatter in Figure~\ref{fig:EffectiveHamiltonian4x4x4} 
is a bit larger, maybe $\pm 0.13$ eV in place of $\pm0.008$,
and with 896 linkages leads to the same estimate.}

\SAVE{In an old version of the figure, ``x-bbond'' is a naive change of the pair;
``bc''-bond is the energies of a (1/2, 1/2, 3/2) neighbor pair.
The latter (WC says elsewhere) is computed with Ca filled centers in
the background clusters.}

The fitted Hamiltonian can be further checked for
transferability to a larger super cell. 
Figure \ref{fig:EffectiveHamiltonian4x4x4} shows a comparison 
between the actual atomic energies of a random configuration 
in a \supcell{4}{4}{4} with the predictions from
the fitted cluster-pair Hamiltonian
It too shows very high correlation between the energies which 
suggests that the effective Hamiltonian is doing a decent 
job of capturing the essence of the interactions.

\subsection{Cluster Hamiltonian with pair potentials}
\label{sec:results-pair-pot}

\TODO{CLH: Hide part!  Need to process the $b$-bond and $c$-bond interaction 
matrices from EOPP into text.  Need to get SVD for $b$-bond, I don't 
find the it in results sent by Woosong.}

\begin{table}
  \centering
  \begin{tabular}{c |  c  c  c  c |}
       &$\dir{X}{+}{r}$ & $\dir{X}{-}{r}$& $\dir{X}{+}{l}$& $\dir{X}{-}{l}$\\
\hline
$\dir{X}{+}{r}$ & -0.0055  &   0.0008  &   0.0013  &  0.0051 \\ 
$\dir{Y}{+}{r}$ & -0.0069 &    0.0012 &  -0.0020  &  0.0025 \\
$\dir{Z}{+}{r}$ &  -0.0021 &   -0.0012 &   0.0044  &  0.0025 \\
$\dir{X}{-}{r}$ &  -0.0054 &    0.0030 &  -0.0024  &  0.0013 \\
$\dir{Y}{-}{r}$ &   0.0027 &   -0.0031 &   0.0017  &  0.0044 \\
$\dir{Z}{-}{r}$ &  -0.0027 &    0.0008 &   0.0017  & -0.0020 \\
$\dir{X}{+}{l}$ &  -0.0072 &    0.0063 &   0.0030  &  0.0008 \\
$\dir{Y}{+}{l}$ &  -0.0137 &    0.0144 &   0.0008  & -0.0012 \\
$\dir{Z}{+}{l}$ &  -0.0158 &    0.0144 &  -0.0031  &  0.0012 \\
$\dir{X}{-}{l}$ &   0.0126 &   -0.0072 &  -0.0054  & -0.0055 \\
$\dir{Y}{-}{l}$ &   0.0220   & -0.0158 &  -0.0027  & -0.0021 \\
$\dir{Z}{-}{l}$ &   0.0220  &  -0.0137 &   0.0027  & -0.0069 \\
\hline
  \end{tabular}
  \caption{Interaction matrix for two clusters separated
by $[1/2,1/2,1/2]$ (=$c$-linkage), using pair potentials (in meV),
using the same conventions as Table~\ref{tab:cbondInteraction}.}
\label{tab:cbondInteraction-pp}
\end{table}

All the fits mentioned till now were derived from EAM potentials.
In this section, we have redone them using fitted (EOPP) pair potentials.
It is not {\it a priori} obvious whether pair potentials should
be adequate for our problem.  The Cd$_{20}$ cage atoms make
large displacements in response to rotations of the tetrahedron,
which might have the same cause as the anomalous $c/a$ ratio in hcp elemental
cadmium, which (at least is the similar element Zn) is poorly captured
by the EOPP potentials~\cite{EOPP}.  The response of cage atoms must 
be the most important determinant of the tetrahedron's elastic interaction 
with its surroundings.

\begin{table}
  \centering
  \begin{tabular}{c | c |   c c c c }
  \hline
 $\sigma$  & irrep & 
    $\dir{X}{+}{r}$& $\dir{X}{-}{l}$& $\dir{X}{+}{l}$& $\dir{X}{-}{r}$\\
  \hline
$89.95$               & A & $-0.1054$ & $-0.0094$ & $-0.3398$ & $0.4546$ \\
$23.83$               & E & $0.2102$ & $0.3825$ & $0.4328$ & $0.5374$  \\
                      &   & $-150.8^\circ$ & $107.9^\circ$ & $-26.9^\circ$ & $141.9^\circ$ \\
$18.43$               & A & $0.4158$ & $0.0999$ & $-0.3514$ & $-0.1643$ \\
$5.612$               & E & $0.6982$ & $0.3914$ & $0.0286$ & $0.1588$  \\
                      &   & $178.0^\circ$ & $70.2^\circ$ & $160.5^\circ$ & $146.9^\circ$ \\
$2.706$               & E & $0.2354$ & $0.4099$ & $0.6458$ & $0.1615$  \\
                      &   & $-62.1^\circ$ & $24.2^\circ$ & $-6.7^\circ$ & $30.3^\circ$ \\
$2.221$               & A & $-0.2569$ & $0.4898$ & $-0.1050$ & $-0.1279$ \\
$3.075\times 10^{-2}$ & E & $0.2822$ & $0.4462$ & $0.2478$ & $0.5714$  \\
                      &   & $-163.5^\circ$ & $-121.0^\circ$ & $55.1^\circ$ & $48.3^\circ$ \\
  \hline
  \end{tabular}
  \caption{ Pair-potential results:
singular values and right singular vectors for $c$-linkage,
in same format as table~\ref{tab:cbondSVD}
}
\label{tab:cbondSVD-pp}
\end{table}

\begin{table}
  \centering
  \begin{tabular}{c | c | c c c c }
\hline
$\sigma$ (meV)  & irrep & $\dir{X}{+}{r}$&$\dir{Y}{+}{r}$&$\dir{Z}{+}{r}$&$\dir{Z}{-}{l}$\\
    \hline
58.15 & $A_2$  & 0.4469 &      $-0.2241$ &     0    &    0    \\Z
8.421  & $A_1$  & 0.1003 &      $-0.1868$ & 0.5309 & $-0.3581$ \\
3.699  & $B_2$  & $-0.2619$ &      0.4259 &     0    &    0    \\
3.487  & $B_1$  &  0.0172   &   $-0.1127$ & 0.2854   &  0.6266 \\
2.369  & $A_1$  & $-0.3560$ &      0.0052 & 0.3651   &  0.3364   \\
2.098 & $B_1$  & 0.3177    &      0.2864 & 0.3579   & $-0.0774$ \\
1.051  & $A_1$  & $-0.1728$ &      0.3630 & 0.0383   &  0.4187   \\
$8.957 \times 10^{-1}$ & $A_2$   & 0.2241   &      0.4469 &     0    &     0     \\
$2.749 \times 10^{-1}$ & $B_1$  & 0.3844    &   $-0.2561$ & $-0.2704$ & 0.0100   \\
$1.925 \times 10^{-1}$ & $B_1$  & $-0.0314$ &   $-0.2995$ & 0.4662   & $-0.3183$ \\
$6.157 \times 10^{-2}$ & $B_2$  & 0.4259    &      0.2619 &     0    &    0      \\
\hline
\end{tabular}
\caption{Singular values $\sigma$ and right singular vectors for $b$-linkage,
as derived from pair potentials, using the same conventions as Table ~\ref{tab:bbondSVD}
}
\label{tab:bbondSVD-pp}
\end{table}

The result is that the interactions derived from pair potentials are remarkably 
consistent with those from EAM.
A sample of this is given by the singular value decomposition of the $c$-bond
interaction (Table~\ref{tab:cbondSVD-pp}), in which the top three
singular vectors show great agreement.

This gives some reassurance as to the independence of our results
from the specific potentials used.
It may help justify the adoption of fitted pair potentials for related 
compositions (in particular ScAn$_6$) for which EAM potentials
are not available, subject to the caution that the EOPP potentials
must be re-fitted if the lattice is compressed or expanded.

\section{Results: continuous orientations}
\label{sec:results-continuous}

Till now, the discussion in this article was limited to a set of discrete 
reference orientations, which are determined by the local minima of
the one-cluster potential.
In fact, that is not a necessary condition for our analysis: our
method of relaxation with the rotation constraint 
(Section~\ref{sec:constrained-relax})
lets us evaluate the the effective Hamiltonian for 
{\it any} set of orientations, whether stable or not.

To proceed, we define some discrete set $\{ \ulo_\alpha\}$ of 
$m$ orientations that samples the continuous space.
We adopt the  uniform background set-up (Sec.~\ref{sec:uniform-background},
letting two interacting clusters run through all combinations
of $(\ulo_\alpha,\ulo_\beta)$ while keepting the other clusters
in fixed orientations.  Once again, we use the singular-value decomposition 
(SVD) \eqref{eq:SVD-form} as the fitting procedure to obtain the single-cluster
and pair terms $U(\ulO)$ and $V(\ulO,\ulO')$.
Since there are now thousands of distinct pair combinations, the SVD 
becomes a necessity in the case of continuous orientations, rather than
an option as it was with the discrete version of the interaction.

The continuous formulation offers multiple opportunities.
First, we can now map out the single-cluster term, separated
from the interaction term.  
This opens up the possibility of detecting metastable, higher 
energy minima, of finding the barriers between discrete wells.
and of simulating temperatures so high that large deviations from
the optimal orientations are typical  in the ensemble.

Secondly, we obtain the cluster pair interaction 
valid for any combination of continuously variable orientations.
even when the interaction term is 
so strong as to destabilize the local well for certain
combinations of orientations (or strong enough to significantly
displace the orientations from the reference orientation).
Finally, going past the SVD this naturally pushes us
to a further step of simplification in representing the
coupling of each cluster, namely orientational harmonics,
and these offer the possibility to unify the basis of
functions $g(\Omega)$ used in representing all the 
different interactions of a cluster.

In this section, we start off by laying out the quaternion-based
mathematical framework needed to describe rotations 
(Sec~\ref{sec:quaternions}).  Then we carry out two forms 
of continuous fit.
First (Sec.~\ref{sec:3-circle}), we limit the rotations of both 
tetrahedra to a single rotation axis.
Second (Sec.~\ref{sec:all-quaternions}) we endeavor to sample {\it all}
rotations.    In either analysis, we do find that 
just the first two or three $\sigma_\mu$ values matter, and
the singular vectors can be interpolated by smooth functions,
thus vindicating the motivation of the singular value analysis
of the potentials.

\subsection{``One-dimensional'' rotations}
\label{sec:3-circle}

\begin{figure}[h]
  \begin{center}
    \includegraphics[width=0.92\columnwidth]{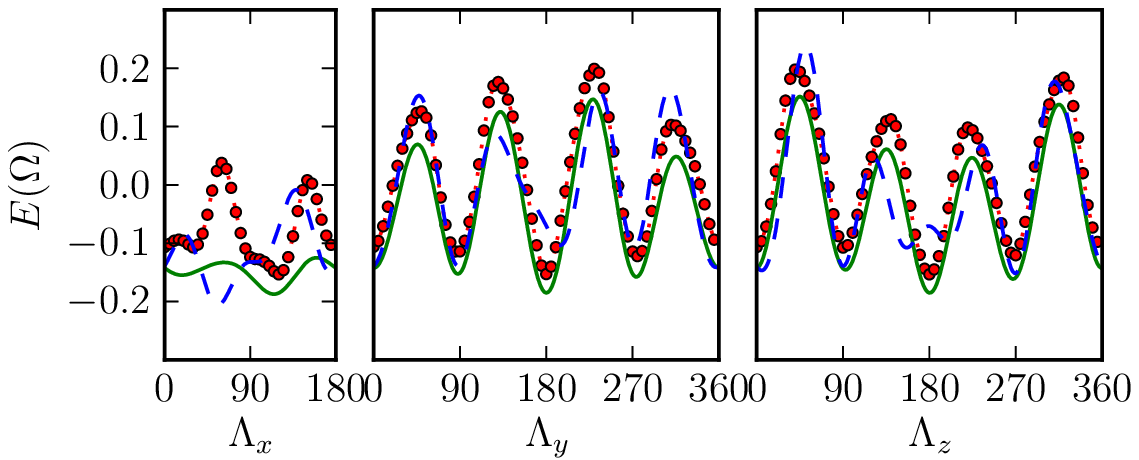}
     \par
    \includegraphics[width=0.96\columnwidth]{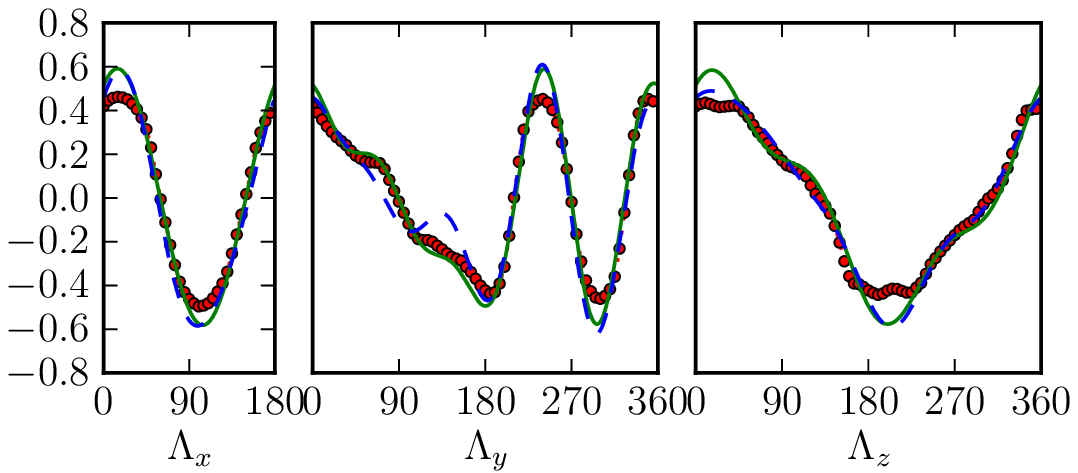}
  \end{center}
  \caption{Fitted potentials for a cluster pair separated by $[0,0,1]$,
displayed along three circular paths in rotation space
parameterized by rotation angle $\Lambda,\Lambda_y,\Lambda_x$ respectively.
(a). single-body potential term $\UC(\Lambda_k)$ along the three
circles (the $\Lambda_x$ plot has period 180$^\circ$).
(b). Singular vector $g(\ulO)$ for the dominant singular value.
(The singular vector for the second cluster is related by symmetry to that
of the first.)
Different lines indicate the following 
databases: 
(i) Dots [red online] = actual measured energies, sampled every 5$^\circ$ 
for a total of 180 points; this was the database for the ``three-circle''
fit; the smooth interpolated version of that, using hyperspherical harmonics,
closely follows the data and is not shown.
(ii) solid line [green online]
= hyperspherical harmonic fits using Polytope 600.
(300 orientations placed on the vertices of
the regular polytope 600 in rotation space, fitted by SVD, 
and then interpolated using hyperspherical harmonics).
A random fit ``R150'' was practically identical to ``P600''.
(iii) The dashed line  [blue online] is ``R30'', a database
of 30 randomly chosen orientations, symmetrized
according to the two mirror planes of the [0,0,1] cluster linkage
that do not swap clusters, for a total database of 120 points.
}
\label{fig:ContinuousFit} 
\end{figure}

\TODOWC{1 WC confirmed that the continuous fits were 
NOT averaged over all backgrounds.  What backgrounds WERE
used?  (i) For the one-dimensional rotations, this matters
because it determines the ``one-body'' contribution shown in
Figure~\ref{fig:ContinuousFit}(a)
(ii) For the polytope sampling, WC has all three backgrounds for the (001) bonds,
[note in draft of 9/11/12].
WC commented ``There are three distinct types of backgrounds where the background
orientations lie along x, y, or z. 
But, for example, $\dir{X}{+}{r}$ backgrounds and $\dir{Z}{+}{r}$
backgrounds give very close singular interactions.''
CLH worries: it does not work the same way for the $c-$axis.  That is,
$\dir{X}{+}{r}$ and permutations $(XYZ)$ are all similar, but
$\dir{X}{-}{l}$ would be quite different.
}
\DONEWC{$\dir{X}{+}{r}$ backgrounds were used for all data presented here, except P600z data below.}
\TODOWC{2 This figure was inserted for internal use only, file
name
``xz\_hyperspherical\_bfit\_x.eps'', was this emailed to CLH
after chatting on Sept 12?
This shows singular values as in Figure~\ref{fig:ContinuousFit}(d).  
The point of the figure
is to confirm it was OK to use only one background for the P600
data, as indicated by the close agreement of the line
marked ``P600'' (also drawn in Figure~\ref{fig:ContinuousFit}(d))
and the line marked ``P600z'' (added here).
But I still don't see exactly what are the two lines.
Is ``P600'' based on only $X$ or on three orientations, and if
so which three?  Is ``P600z'' based only on a Z background?
\includegraphics[width=0.9\columnwidth]{xz_hyperspherical_bfit_x.eps}
}
\DONEWC{This is a good point, I have not tried different backgrounds with c-bonds. P600 should be
labeled P600x to be consistent with P600z. P600(x) is in $\dir{X}{+}{r}$ background, and P600z in
$\dir{Z}{+}{r}$.}

The simplest continuous sub-space of the manifold of orientations consists of
all rotations around one axis, parametrized by a single Euler angle.
\SAVE{(This is convenient for plotting functions of the otherwise
three-dimensional manifold of rotations.}
We choose this axis such that the
rotations connect (at least) two of the known optimal orientations.
Specifically, we choose a zero orientation of $\dir{X}{+}{r}$ 
so that rotation around the $x$-axis passes through $\dir{X}{+}{r}$ 
to $\dir{X}{+}{l}$, $\dir{X}{-}{r}$, and $\dir{X}{-}{l}$. 
(a rotation by exactly $\pi$ takes $\dir{X}{+}{r}$ to $\dir{X}{-}{r}$ 
or $\dir{X}{+}{l}$ to $\dir{X}{-}{l}$.) 
Rotations  around $y$ and $z$ axis are also used; we will call the
respective rotation angles $\Lambda_x$, $\Lambda_y$, and $\Lambda_z$.
The sampling points are spaced by 5$^\circ$ along each of the three circles.
These three data sets are combined into a single one, which we call the
``three-circle'' point set, with a total of 36+72+72=180 sample points 
(the $\Lambda_x$ rotation runs only to $180^\circ$ since it repeats 
after that.)

The first result of the fit (before any SVD analysis) is the
single-body potential, a byproduct of the processing mandated in 
Sec.~\ref{sec:redundancies}.  
(Recall that such information cannot be obtained
from the discrete analysis of Sec.~\ref{sec:results-discrete}.)
The results are shown in Figure~\ref{fig:ContinuousFit}(a).

Each deep well is one of the optimal directions. Angle $\Lambda_{x,y,z}=0$ is, 
by definition, the orientation $\dir{X}{+}{r}$.  Rotating around the $x$ axis, 
we encounter $\dir{X}{-}{l}$ at $90^\circ$; also, at 
$\Lambda_x\approx 30^\circ$ and $\Lambda-x\approx 120^\circ$,
we meet $\dir{X}{+}{l}$ and $\dir{X}{-}{r}$, respectively.
At either $\Lambda_y=180^\circ$ or $\Lambda_z=180^\circ$,
we meet $\dir{X}{-}{r}$; at $\Lambda_y=\pm 90^\circ$ we
meet $\dir{Z}{\pm}{r}$, while at 
$\Lambda_z=\pm 90^\circ$ we meet $\dir{Y}{\pm}{r}$.
The double well at $\Lambda_x=105^\circ \pm 15^\circ$ 
is responsible for the $\pm 15^\circ$ rotation of the
tetrahedra in optimal states (of these, the well
at $\Lambda_x=90^\circ$ actually appears to be
destabilized by the uniform background.)

In principle, the single-body output has contributions
(as in the discrete case) from the pair interactions of the 
background, as well as from the true single-cluster term $U(\ulO)$;
however the latter contribution is much larger. The background 
contribution merely creates slight energy offsets, visible 
in the figure, between wells which ought to be symmetry-equivalent:
e.g. along the $\Lambda_x$ circle,
$\dir{X}{-}{r}$ appears to be lower in energy than $\dir{X}{+}{r}$,
by $\sim 0.04$ eV.   
(We would eliminate the background contribution if we averaged over
all possible backgrounds, as was done in the discrete analysis of
Sec.~\ref{sec:results-discrete}, but that was not carried out
in our treatment of the continuous rotations.)

Next comes the singular vector analysis, with the singular vectors
normalized according to Eq.~\eqref{eq:continuous-normalize}.
The resulting singular values are included in
Table~\ref{tab:ContinuousSingularValues}.
The overall magnitudes are comparable to the discrete result, and
some differences can be explained away because the three-circle 
data set is not even approximately uniform: e.g., as it starts
from $\dir{X}{+}{r}$, it over-represents the four $X$ flavors of 
discrete orientation.  (Since those ones have the strongest pair interactions, 
according to Table~\ref{tab:bbondInteraction}, it is not surprising
that the dominant singular value for the $b$-linkage for the C3 data
set (Table~\ref{tab:ContinuousSingularValues}) comes out 20\% larger 
than the corresponding singular value for optimal orientations
in Table \ref{tab:bbondSVD}.)
Note also that the third singular valus is not so well separated here
from the second one, as was the case with the discrete orientations
(Sec.~\ref{sec:results-SVD}).

 \begin{table}
   \centering
   \begin{tabular}{c || c c c | c c c }
     Singular value    & \multicolumn{3}{c}{c-linkage} & \multicolumn{3}{c}{b-linkage} \\
       & C3 & P600 & R30 & C3 & P600 & R30 \\
     \hline
     \hline
     $\sigma_1$ & 
          $80.28$ & $86.04$ & $68.06$ &
          $95.20$ & $76.52$ & $67.86$ \\ 
     $\sigma_2$ & 
          $51.64$ & $50.87$ & $43.04$ &
          $16.17$ & $25.80$ & $18.40$ \\ 
     $\sigma_3$ & 
          $28.70$ & $48.25$ & $31.93$ &
          $11.00$ & $6.648$ & $6.416$ \\
   \end{tabular}
   \caption{First three singular values from three continuous data sets (in meV).
Here ``C3'' is the three-circle data set (Sec.~\ref{sec:3-circle}),
while ``P600''  and ``R30'' are the polytope 600 and the random 30 data sets
(Sec.~\ref{sec:all-quaternions}).}
   \label{tab:ContinuousSingularValues}
 \end{table}

In conclusion, the three-circle data set gave a decent indication
of how smoothly the potential varies between the relevant discrete orientations,
and it goes along with a convenient way of plotting singular vectors
along cuts in this three-dimensional parameter space.
However, it completely fails when we try to transfer to orientations that
are not near to those three circles in orientation space.
Thus, this approach does not suffice to provide a parameterized representation 
of the complete functional form 
for {\it all} possible orientations, as we would need to use this in a simulation
with continuous angles.  
\SAVE{(Woosong EMAIL 8/30/12)
The ``slices fit'' (using only the three circles of single-angle rotations) 
does a terrible job when we look at rotations starting from other orientations.
It fluctuates wildly, for example, when we  rotating $\dir{Y}{+}{r}$ 
around the $x$ axis.}

\SAVE{
Should the three-circle data set become sufficient,
if we fit to the minimal set of hyperspherical harmonics?
The answer is no: since we started from a 
particular orientation of cluster, which has a particular relationship
to the inter-cluster direction, this does not sample orientation
space in a symmetrical fashion.}

\SAVE{
There IS an extension of the three-circle idea that might work,
but which we have not tried. 
Generate a list of orientations by taking ALL 12 ideal optimal
orientations, and rotating each of them about all three axes with a
relatively coarse angle spacing (say, by 15$^\circ$ increments).
Owing to the twofold symmetry, 
the rotation axis aligned with the twofold symmetry axis 
of this orientation must only 
be taken to $180^\circ$, while each of the other two circles actually
includes four other ideal orientations so they must only be
taken to 90$^\circ$.  Thus the resulting list has $12(1+11+5+5)$
or 264 orientations in it.  If all combinations of these are tried,
and the singular vectors are represented using hyperspherical harmonics,
that might be sufficient to define the orientational potential.}

\TODOWC{4a (About figure~\ref{fig:ContinuousFit}:)
We agreed it would look better not to use $\times$ but instead to
use maybe dots that are connected together, for (a) and (b).  Did WC send
those already?  I also wonder now, is the curve in (c) called ``Data''
the same exact curve as plotted in (b)?  If so, I guess that panel (b)
ought to be omitted -- do you agree?  Also, I wonder if the key for
that curve in (c) could be labeled something other than ``Data''; I would
call this the ``one-dimensional'' or ``single circle'' data.
Well, actually I don't like showing little keys in the figure at all,
and I don't like relying on color contrast alone;  I prefer to draw a
label next to the line or pointing to it. (I can add that using xfig).}
\DONEWC{They are the same curves, and I agree (b) should be omitted.
The legends have been removed on the new figures}
\TODOWC{4b Now I'm thinking that, I'd like to see a plot of the single
body energies from P600, just to see if it looks like (a), but not
necessarily to put in the paper.}

\begin{figure}
\label{fig:HypersphericalFit-c}
  \begin{center}
    \includegraphics[width=0.96\columnwidth]{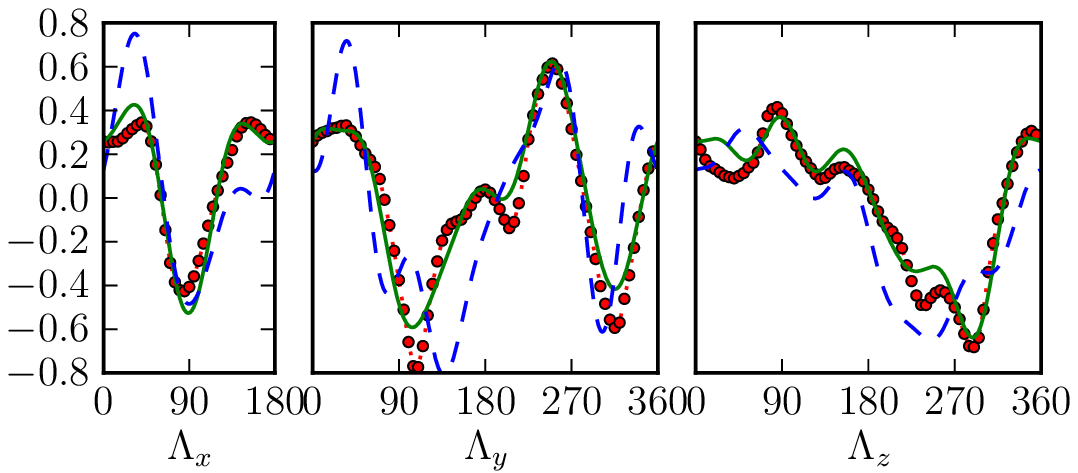}
     \par
    \includegraphics[width=0.96\columnwidth]{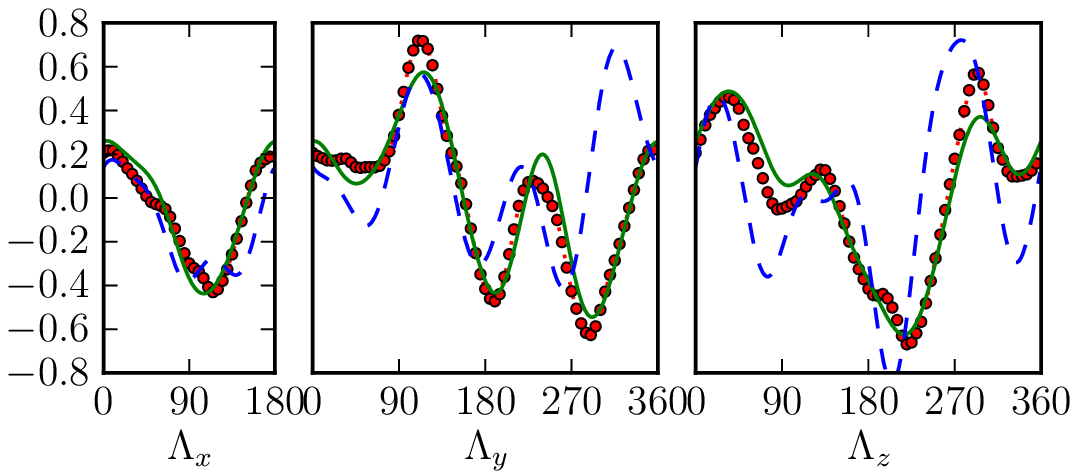}
     \par
    \includegraphics[width=0.96\columnwidth]{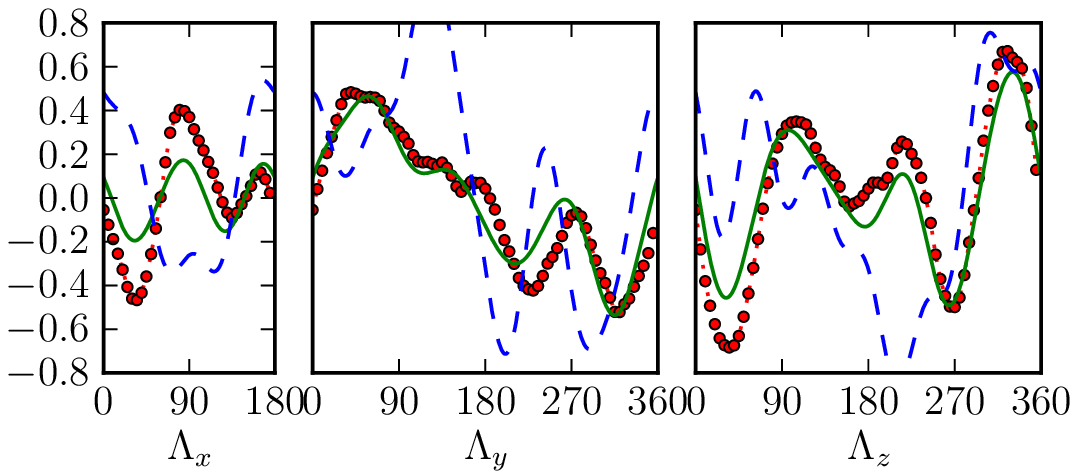}
  \end{center}
\caption{Same as Figure \ref{fig:ContinuousFit}(b), but for
clusters related by a ``$c$'' type separation. Two additional singular 
vectors are illustrated, as they are comparable in strength to
the leading ones.
Note that this fit is less successful, and in places the R30 fit fails.
(a) Dominant singular vector (b) Second singular vector (c) Third singular vector.
(The single-body energy plot is very similar to Figure \ref{fig:ContinuousFit}(a)
and is not shown here.)
}
\end{figure}

\TODOWC{5a TASK/Discuss The hyperspherical singular vector files 
in Figure~\ref{fig:ContinuousFit} (c), also in the next figure,
now have just
two kinds of plain line in them, which are blue and green.  Could we instead
make one of them b dashed (while keeping different colors), perhaps the
random 30? (I also wonder if it is necessary to connect the data points
with a line, but I will defer to your choice on that one.)
I would then remove the legend identifying the lines.
Also, recall the ``...cfit\_x3'' plot had the vertical axis reversed
for the data curve; and the axis labels are overlapping.
Also, these panels are very high, so when I have three of them
stacked (as in Figure~\ref{fig:HypersphericalFit-c}) it takes
a lot of space.  Do you think they could be made shorter in that
direction, and still show the needed information?  Do you
think it is valuable to show all three singular values in
Figure~\ref{fig:HypersphericalFit-c}? }
\DONEWC{Redrawn figures. (WC note of 9/12/12).
(But were the new figures sent to CLH?)}
\TODOWC{5b CLH wondered if R150 still looks like P600 for
the c-bond.
WC: R150 behaves similarly well as P600. But, as much as P600 is doesn't get
the shapes perfectly, R150 doesn't either. (Maybe I'm in need of more
hyperspherical modes for the fit) See the separate 
file ``cbond\_r150.png'' black dotted line is R150 in the figure.}

\subsection{Mathematical handling of continuous rotations}
\label{sec:quaternions}

The most uniform way to parametrize rotations in three dimensions is by 
four-component unit {\it quaternions}: for a 
rotation of angle $\theta$ about axis $\hat{\theta}$, the first component
is $\cos(\theta/2)$ and the other three are $\sin(\theta/2)\hat{\theta}$.
Thus the quaternions map out a hypersphere, which corresponds 2-to-1 with
rotations (since antipodal quaternions represent the same rotation).
The quaternion components are related to the three Euler angles 
as follows:
   \begin{subequations}
   \label{eq:quaternion-Euler}
   \begin{eqnarray}
    q_0 &=&\cos\alpha  \\
    q_1 &=&\sin\alpha\sin\theta\cos\phi  \\
    q_2 &=&\sin\alpha\sin\theta\sin\phi  \\
    q_3 &=&\sin\alpha\cos\theta 
   \end{eqnarray}
   \end{subequations}
Here $2\alpha$ is the total rotation angle, and
$(q_1,q_2,q_3)$ equals $\sin\alpha$ times the unit vector
of the rotation axis.

The proper measure in rotation space is uniform on the unit
hypersphere parametrized by \eqref{eq:quaternion-Euler}.
Thus, the normalization convention used for functions of rotation space is
   \begin{equation}
   \int d^3\ulo \equiv (\int_0^{2\pi} \sin^2\alpha d\alpha)
   (\int_0^\pi \sin\theta d\theta)(\int _0^\pi d\phi).
  \label{eq:rotation-int-normalization}
  \end{equation}
In particular, the total volume of $\ulo$-space is $(2\pi)^2$.
Hence, we normalize a discrete singular vector $(g_1,g_2, ...)$ by
  \begin{equation}
   \sum_{k=1^m} |g_k|^2 \equiv \frac{m}{\pi^2}.
  \label{eq:continuous-normalize}
  \end{equation}
If our singular vector is the sampling of a normalized
continuous mode $u(\ulo)$ and if the sampling points are
uniformly distributed, this will be equivalent to the 
normalization of the continuous mode.  With such a normalization,
singular values obtained from different constructions 
ought to have similar magnitudes, as is borne out in
Table~\ref{tab:ContinuousSingularValues}.

\subsubsection{Hyperspherical harmonics}
\label{sec:hyperspherical}

We assume the interaction (or singular vector) is a 
smooth function in rotation space. The standard way
to parametrize such a function with a small number of fitting parameters, 
is a series expansion using some basis of orthogonal functions,
ideally tailored to the symmetry of the space. 
When our data, sampled at necessarily sparse points in rotation space,
is expressed in this basis, it can be thought of as simply
an elaborate kind of interpolation.

The natural basis for the 3-sphere is the 
{\it hyperspherical harmonics}, 
analogous to expanding  in spherical harmonics on a 2-sphere.
(These are also commonly used in the theory of textures in
materials science~\cite{hyperspherical}.)
We adopt the definitions and normalizations for real hyperspherical functions
from Eq. (6) of Ref.~\onlinecite{hyperspherical}.
These carry three indices: $N$, the total ``hyperangular momentum'', with $L$ and $M$ 
to label different functions within the same representation.
The real hyperspherical harmonics, in terms of the three Euler angles
$\alpha, \theta, \phi$, are then given  by
  \begin{subequations}
  \begin{eqnarray}
   Z^{MC}_{NL}(\ulo) &=&  z_{NLM}
   C^{L+1}_{N-L}(\cos\alpha) P^M_L(\cos\theta) \cos M\phi;  \\
   Z^{MS}_{NL}(\ulo) &=&  z_{NLM}
   C^{L+1}_{N-L}(\cos\alpha) P^M_L(\cos\theta) \sin M\phi. 
  \end{eqnarray}
  \end{subequations}
where $C^{L+1}_{N-L}(\cos\alpha)$ is a 
Gegenbauer polynomial and $P^M_L(\cos\theta)$ is an associated Legendre function.
These are orthonormalized with respect to the measure of Eq.~\eqref{eq:rotation-int-normalization},
and the normalization constant is
  \begin{eqnarray}
   z_{NLM} &=& (-1)^{L+M} \frac{2^L L!}{\pi}  \Bigg[ (2L+1) 
       \nonumber \\ &&\frac{(L-M)!}{(L+M)!}  
       \frac{(N+1)(N-L)!}{(N+L+1)!} \Bigg]^{1/2}.  
  \end{eqnarray}

\subsection{Fits from sampling all orientations}
\label{sec:all-quaternions}

It is necessary to devise some roughly uniform way to sample rotation space.
We have tried two ways.  First, we choose $m$ quaternions uniformly spaced by 
placing them  on one of the regular polytopes of icosahedral symmetry, the higher-dimensional
analog of an icosahedron.  However, it appears the polytope is inefficient because 
it has too much symmetry, and some places in rotation space are far from any point
of the polytope.  Our second approach is to select a random list of orientations  $\{ \ulo_k \}$ 
and then apply the $mm$ symmetries around the $[0,0,1]$ bond,
i.e. those which preserve the two cluster positions.
This turned out to work much better.

\TODOWC{7 Query.
WC's reply on 9/11/12: ``The symmetries applied are not relevant to the type of bond. 
I'm applying (proper) permutation symmetries of each tetrahedron that corresponds to rotations.
Because the tetrahedron rotated with respect to its own permutation symmetry should
interact exactly same with any neighbors, these are independent of the bond type.''
If that is accurate, then my previous statement about the $mm$ symmetry is
in error.  My understanding is that certain pairs don't need to be computed over again
because they are equivalent by symmetry, and that (in this particular computation)
you took advantage of that symmetry.  Obviously, the symmetries are different in the
case of the $b$-bond and the $c$-bond...  There is a separate but related issue I 
newly realized (see TODOWC 1), which is about which symmetry operations were
applied to the BACKGROUNDS used for these fits.}
\DONEWC{TODOWC 1: for discrete orientations, (pressurized runs) $mm$ symmetry including
that of the background are used to not do precisely equivalent configurations. For continuous
orientations, only single tetrahedron symmetries are used~(i.e. $\dir{X}{\pm}{r/l}$ invariant
under $180^\circ$ rotation around $x$-axis.) thus background is irrelevant.}
\TODO{CLH EXPLAIN THE FIGURE here}

\SAVE{We can check the separation of the polytope, with some estimtes.
If we imagine space is divided into $m$ regular, small tetrahedra,
then $2 \pi^2 / m$ is the mean hyper-solid-angle that each point ``covers'',
and the volume of the small tetrahedron.
For Polytopes 120 ($m=120$) or 600 ($m=600$)so 
we get roughly $31.4^\circ$ and $18.3^\circ$ as the angular distance between each point. 
(Direct evaluation of the polytope points in orientation space gives 
$72.0^\circ$ and $31.0^\circ$ for the respective angles between nearest points, 
which confirms that the numbers are in the ballpark.)}

Applying the SVD yields one dominant singular value,
along with a corresponding dominant singular vector 
$g_{\mu\alpha}$
as shown in Figure~\ref{fig:ContinuousFit} (b,c).
This represents the leading mode of the two-body interaction,
and looks like sampling a largely sinusoidal 
function of $\ulo_\alpha$ [Figure~\ref{fig:ContinuousFit}(b)].
The $y$-axis shows a noticeably non-sinusoidal profile;
\LATER{MM Do the modulations in this curve correspond to the
glitches  visible in Figure~\ref{fig:ContinuousFit}(a)?
CLH saw these as sharper
peaks added to the shoulders of larger smooth peaks.}

Fitting the coefficients of hyperspherical harmonic expansion 
to the most important singular vector yields,
for the case of the $[0,0,1]$ (or ``$b$'') interaction,
coefficients in the $N=6$ and $N=8$ hyperspherical harmonics. 
Further harmonics were not needed and (when included) seemed
to reflect sampling arbitrariness and not improve the fit.
For the $[1/2,1/2,1/2]$ (i.e. ``$c$'') interaction, 
we additionally needed $N=12$ in order to get a decent fit.

On the other hand, the hyperspherical fits of the $[1/2,1/2,1/2]$ 
(``$c$'') cluster pairs require
$N=12$ hyperspherical harmonics in addition to the $N=6$ and $N=8$ 
used for the $b$-interactions, and still this is not so good a fit. 
Furthermore, the second and third largest singular values are 
non-negligible in the case of the $c$ interaction, as shown in
the figure.  We conjecture that the $c$ interaction is the most
complicated simply because it is the shortest. 
\SAVE{But even in that direction, the Zn cage atoms from one cluster 
are three atom steps away frome those of the other cluster.}

\TODO{FIGURE:
The singular vectors plots for 2nd and 3rd
singular vectors also plotted - cfit\_x2 and cfit\_x3. cfit\_x3 has the P600
with flipped sign which WC needs to manually correct for, has that
been done?}

\TODO{CLH: WC originally wanted to
understand the constraints on the hyperspherical coefficients
corresponding to tetrahedral, cubic, (and icosahedral?) symmetries
in this problem.}

Figure \ref{fig:ContinuousFit}(c) shows how the fitted singular vectors 
compare against the singular vector from slices measurements. 
The fits from polytope-600 and the three-circle data set both give 
reasonable approximations of the actual data. 
In contrast, polytope-120 (not shown) fails completely,
which can be readily understood from the fact that the orientations 
in polytope-120 are too spaced too far apart ($72^\circ$  from
each other), and furthermore the cuts shown in the figures
do not even have to go through the sampled orientations.

\SAVE{WC noted summer 2012:
There ought to be a smaller set of points (than Polytope 600)
that we could use to find the form faster.  
Specifically, 
hyperspherical harmonic fits that WC used, with hyper-angular momentum 6 and 8, 
have less than 150 coefficients. Because of the tetrahedral symmetries, 
however, the number of independent coefficients is smaller.
Therefore WC thinks that -- provided that the 6 and 8 hyperspherical harmonics
explain the interaction well -- we should only need (at most) $\sim 150$
independent orientations to fit nicely. Indeed,
this this seems to be the case with random 150 orientations.}

\section{Orientational Orderings} 
\label{sec:ordering}

\TODO{CLH: read Brommer thesis,  describe
enumerated ground state~\cite{brommer-thesis}.}

A great advantage of the fitted Hamiltonian approach is
that one may discover the optimal structures for systems that
were not included in the fitting database (and which could not
have been included, because they are too big).  In this spirit,
we take the discrete effective Hamiltonian from 
Section ~\ref{sec:results-discrete} and find what is predicted
for the ordering pattern at low temperatures.  
We used Monte Carlo annealing  in supercells
to discover the ground state.  

Watanuki et al.~\cite{watanuki2006} discovered pressure-induced 
phase transitions in the CdYb$_6$ cubic crystal.
Therefore, we extend our studies to 
nonzero pressure, so as to make contact with experiments that
show different ordered states appearing in a range of
pressures $< 10$GPa.

\subsection{Strain, pressure, and bulk modulus}

To extend our calculations to varying hydrostatic pressure,
we reran calculations of the energies and ground states
constrained to various strain values, separated
by $0.002$ (or by $0.010$ for strains greater than $0.01$).
The corresponding pressure $P$ (at $T=0$) was then evaluated by recalculating
the energy at slightly different cell volumes $V$ and using 
$P\approx (\Delta E)/(\Delta V)$ .  
The pressure/strain relationship suggests that 
changes in the lattice constant at the $T=0$ transitions are
quite small. 

Note that the ``strain 0'' results, reported in previous sections, 
used a cell constrained to an a priori lattice constant.
The present calculation shows that (with the EAM potentials we are using) 
they do not exactly not correspond to zero pressure, in fact
we found zero pressure at negative strain.
A possible physical meaning to study even more negative strains
(with negative pressures) is as follows.
In isotructural compounds with the large atom species 
varied, an increase (decrease) in its effective radius is 
appears as a negative (positive) ``chemical pressure'', so
that we might see a similar phase diagram
except for a shift along the $P$ axis.

The bulk modulus grows rather uniformly with pressure:
from about 80 GPa to 300 GPa over the range from
our most negative pressure (strain $-0.020$) to the largest one
(strain $+0.040$).  Specifically it was around  150 GPa at $P=0$ or 200 GPa
at $P\approx$ 1 GP.
We do not know of any experimental measurement of the bulk modulus 
of CaCd$_6$; measurements of the (similar) quasicrystal phase
$i$-CdCa gave a bulk modulus 68.1 GPa at zero pressure~\cite{bulk-mod-iCdCa}.

\SAVE{
\begin{table}
\caption{Pressure-strain relation, showing the
strain, optimum structure, and pressure $P$ for that
optimal structure.
\TODOWC{8b QOK I would hide this table, too.
The transition pressures have been included in
the table of structures.}
\DONEWC{OK}
}
\begin{tabular}{l|lcll}
Strain  &   structure & $P$ (GPa) \\
\hline
$-0.020$     &      ZY  & -2.360\\
\hline
$-0.018$     &  $Z_2$   & $-2.205$ \\
$-0.016$     &   --"--  & $-2.036$ \\
$-0.014$     &   --"--  & $-1.858$ \\
$-0.012$     &   --"--  & $-1.663$ \\
$-0.010$     &   --"--  & $-1.456$ \\
$-0.008$     &   --"--  & $-1.231$ \\
$-0.006$     &   --"--  & $-0.988$  \\
$-0.004$     &   --"--  & $-0.721$  \\
$-0.002$     &   --"--  & $-0.441$  \\
$0.00$     &   --"--    & $-0.138$  \\
$+0.002$     &   --"--  & $0.174$  \\
\hline
$+0.004$     &   $Z_4$  & $0.517$  \\
\hline
$+0.006$     &   $X_4$  & $0.870$ \\
\hline
$+0.008$     &   $XY$   & $1.253$ \\
$0.010$      & --"--    & $1.651$ \\
$0.020$      &  --"--   & $3.982$ \\
$0.030$      &  --"--   & $5.993$ \\
$0.040$      &  --"--  &  $8.800$ \\
\end{tabular}
\label{tab:pressure}
\end{table}
}

\SAVE{The pressure tabulated in hidden Table~\ref{tab:pressure}
before 9/11/12 was $P_{2+u}$, derived from
all the configurations with a cluster pair in one of
the discrete orientations plus a uniform background, 
as were used in fitting the effective potential.  
For each such combination, the pressure was found.
Thus, if we multiply count the configurations related  by
symmetries, there should be $2(12^3)$  pressure 
measurements:
the $2$ comes because both $b$ and $c$ pairs were included.
The stdev of pressure shown in that old table is taken over
all entries in that list.}

\TODO{See emails about the validity of this pressure calculation.}

We also estimated the pressure using all possible
\supcell{2}{2}{1} supercells with a pair of clusters
of all possible orientation combinations, placed in
uniform backgrounds of all possible orientations. 
This was the same database used to construct the cluster
pair potentials.
\SAVE{since we averaged over all orientations of the
pair of clusters, this is like keeping the constant
term in the decomposition.}
These are higher-energy structures, so this is roughly
like an ensemble of infinite temperature (so far as
cluster orientations are concerned).
We found that (i) the pressure was higher by $~\sim 1.5$ GPa,
(ii) at zero pressure, the strain was $-0.008$, i.e. an
increase of the lattice constant which could be interpreted
as an orientational contribution to the thermal expansion.
(iii) the bulk modulus is $\sim 20\%$ higher in the $ZY$ structures
found at small negative strain, to $\sim 10\%$ higher around the
transition to the $ZY$ structure around strain $0.07$, and
roughly unchanged at the highest pressures.

\SAVE{We have not studied pressure effects using the pair potentials,
as the Friedel oscillations in our EOP potentials are critically sensitive 
to the overall valence electron density, which increases with compression.}

\SAVE{In a very old draft:
``The relation of strain to pressure in our data 
(Table~\ref{tab:pressure}) 
is almost linear; the Bulk modulus inferred from this
is about  57.93 GPa.''
This is obsolete. 
Now we see the modulus varies by a factor of two just
within the $Z_2$ phase that appears to be the zero-pressure structure.}

\subsection{Monte Carlo search for ground state in supercells}
\label{sec:MC-grstates}

\SAVE{
For sizes beyond \supcell{2}{2}{1}, exact enumeration became prohibitive 
and one must resort to Monte Carlo simulations.  Woosong did some 
Monte Carlo w/o tempering, slowly annealing to temperatures below $10K$, 
in \supcell{3}{3}{3}, and found irregular structures just slightly higher in energy 
than the \supcell{2}{2}{1};
the \supcell{3}{3}{3} cannot accomodate the optimum pattern.
The \supcell{4}{4}{4} super cell could not be equilibrated by Woosong
or (later) by Marek w/o tempering.
Woosong found state with mean energy per site is not so close 
to the ground state value, 
yet with some clusters having a much better site energy than in
the ground state: this indicates frustration in the problem.}
\SAVE{We could use an inter-cluster
``site energy'' diagnostic, as a clue to which clusters
the frustration is concentrated at.
Woosong noted that in frustrated configurations, the site
energy was very uneven between clusters.
We have not tested this on the \supcell{4}{4}{2} states.}

We performed Monte Carlo (MC) simulations
using the fitted discrete cluster Hamiltonian in a 
\supcell{4}{4}{4} supercell, i.e. 128 clusters. (Sizes over 
$\sim 50$ clusters cannot be equilibrated with plain Monte Carlo
owing to the frustrated interations.)
The lowest energy configuration encountered during a run
was saved and defines the ground state ordering.

\subsubsection{Results: ordered states}

The predicted ground state orientation pattern does change under pressure. 
We have found at least three distinct optimal states as $P$ is varied.
The transitions between them at $T=0$ are necessarily
first-order: since every tetrahedron falls into one of twelve
discrete orientations, there is no way that one pattern can
evolve continuously into a distinct one.

\begin{figure}[h]
  \begin{center}
   \includegraphics[width=1.0\columnwidth]{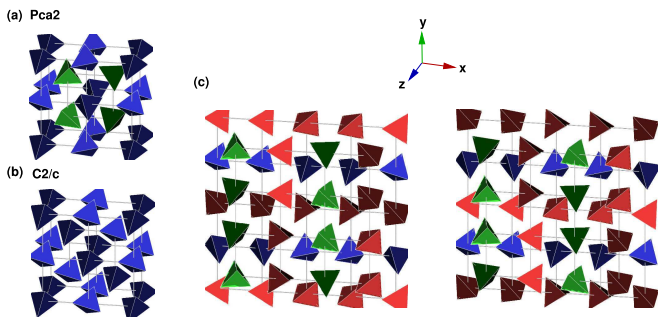}
  \end{center}
  \caption{[Color online] 
Phases of tetrahedron order in
CaCd$_6$ found from simulation at various pressures.
[Online: tetrahedra aligned with $X$, $Y$ or $Z$ directions
are colored red, green, and blue respectively; for each 
direction, a lighter shade 
is used for both orientations in the ``$+$'' sense and darker for
both in the ``$-$'' sense.]
\SAVE{convention confirmed by WC 9/11/12}
They are tagged by a provisional label based on the orientations
shown in this figure, and by their space group
(a) ``$ZY$'' with $2\times 2\times 1$ cell,
Orthorhombic  $Pca2$ [here in the setting $P2ab$].
The bcc lattice corner clusters have the four directions $\dir{Z}{\pm}{r/l}$
while body-center clusters have the four directions $\dir{Y}{\pm}{r/l}$.
(b) ``$Z_2$'' with $2\times 2\times 1$ cell, Monoclinic  $C2/c$; the $2\times 2\times 2$
supercell shown is two unit cells.
Tetrahedra have two orientations, $\dir{Z}{\pm}{r}$.
(c) ``$XYZ$'' with $4\times 2 \times 2$ cell,
CHECK space group in the setting ...
The structure is shown as two slabs, each one lattice constant thick.
\TODO{CLH: need to check all space groups for the
new ground states.}
\TODO{CLH: redo cropping of figures so as not to clip
tetrahedron corners; add correct space group of (c)}
}
\label{fig:orderings}
\end{figure}

\begin{figure}[h]
  \begin{center}
   \includegraphics[width=1.0\columnwidth]{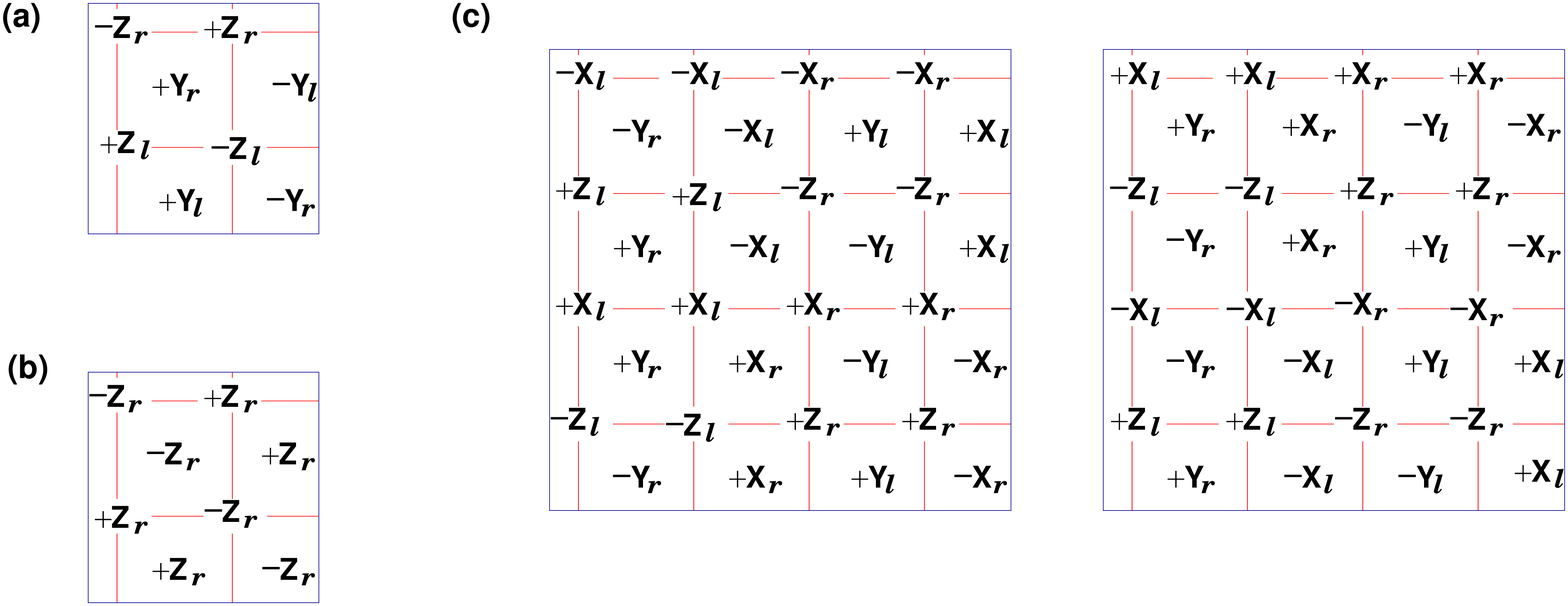}
  \end{center}
  \caption{The same structures shown in Figure ~\ref{fig:orderings},
expressed in terms of the orientation labels defined in 
Sec.~\ref{sec:one-cluster}
  }
\label{fig:orderings-labels}
\end{figure}

The three $T=0$ phases are shown in Figures~\ref{fig:orderings}
andr~\ref{fig:orderings-labels}.
\SAVE{The 2x2x1 enumeration (see Sept 2012 drafts)
contained a ``$Z_4$'' state with all four orientations of types $\dir{Z}{\pm}{r/l}$.
If we do not distinguish the small rotations $r/l$, these have
$\{110\}$ planes with alternating $+Z$ and $-Z$ orientations, 
just like the ``$Z_2$'' state, so these are variations within the same phase.
Also an ``$X_4$'' state contains all four orientations of types 
$\dir{X}{\pm}{r/l}$, and actually fits in a
\supcell{2}{1}{1} cell; evidently, the ordering wavevector
is $(1/2,0,0)$.  Finally, Woosong's table actually had a 
$Y_4$ state with unit cell \supcell{1}{2}{1}.
All the $P>0$ part was simplified when we tried larger supercells}

We saw three patterns in the whole range of pressures
(actually strains):
\begin{itemize}
  \item[(1)]
  at strain $-0.020$, i.e. roughly $P < -2.3$GPa, we saw
a structure we call ``$ZY$'' with a \supcell{2}{2}{1}
unit, thus 8 clusters per cell, which appears to have
orthorhombic space group  $Pca2$
[Figure ~\ref{fig:orderings}(a)]
\TODO{Why did space group finder say it is triclinic
space group No. 1?}
  \item[(2)]
for strains  $-0.018$ to $-0.004$, i.e. pressures $-2.3$ to
$-0.6$ GPa, we found a simpler structure we call $Z_2$
containing just two cluster  orientations which simply form $(110)$ layers.
[Figure ~\ref{fig:orderings}(b)]. The space group
is centered monoclinic $C2/c$ with 4 clusters per cell
(=2 per primitive cell).
\TODO{Space group finder says space group No. 5, which is $C2$.}
  \item[(3)]
For strains $-0.020$ through $0.008$, i.e.
up to a pressure of $\sim 1.4$GPa, we see the complex 
\supcell{4}{4}{2} pattern shown in 
[Figure ~\ref{fig:orderings}(c)], which has a triclinic
(pseudo monoclinic) unit cell; there is a centering
operation within the \supcell{4}{4}{2} cell so there
are 32 clusters per cell.
  \end{itemize}
Finally, for the large strain of $0.010$ (around $P=1.6$ GPa),
we found a less regular \supcell{4}{4}{2}; we do not know if this
is the true ground state.

\TODO{Find and insert the detailed description I wrote of the 4x2x2 phase.}

We will describe the \supcell{4}{3}{2} ordering pattern 
as a stacking of $4\times 4$ layers along the $z$ direction.  
The structure repeats after four of these layers
(i.e. after two lattice constants, as there are layers of cell corners
alternating with layers of body-center sites).
We will use ``flavor" to designate the $X$, $Y$, or $Z$ nature of the orientation,
so that each flavor includes four orientations.
The even layers (cell corners) are of one kind that we call
``$XY$'' layers,
after the orientation flavors found in them.  The odd layers
(body centers) are of another kind we will call ``$XZ$'' layers.

The basic ingredient of a layer is a chain with a pattern of orientations
$AA\bar{A}\bar{A}$ where $\bar{A}$ stands for inversion of $A$, where
$A=X,Y,$ or $Z$.  Every layer is made by stacking such four chains
side-by-side, using two alternating flavors; the second occurrence 
of the same flavor uses the orientations not found in the first one.  
In the $XZ$ layers, the chains run in the $x$ direction and are stacked 
in the $y$ direction. 
Note that, if we draw bonds between neighbors of identical orientations,
this forms a columnar pattern of dimers covering the square lattice.
The $XY$ layers have chains the other way around,
running in the $y$ direction and stacked in the $x$ direction,
but slightly different from the way in the $XZ$ layers.
The slight difference is that here, if we mark the adjacent pairs of the
same orientation, it forms a {\it staggered} dimer pattern.

If we go up two layers, i.e. one lattice constant in the $z$ direction,
we get the same pattern, except all orientations are replaced
by the complementary ones of the same flavor.
Also, the registry between successive layers is such that
(in projection) the $X$ dimers in the $XZ$ layer
cross the $Y$ dimers in the $XY$ layer.

We also did simulations with smaller dimensions of supercell.
Runs with a \supcell{2}{2}{1} cell indeed give the same ground
state as the \supcell{4}{4}{4} whenever that state fits into
the smaller supercell, i.e.  for negative strains;
for positive strains, the \supcell{2}{2}{1} gives higher 
energy states, showing that the ground state cannot fit into
that cell. (Monte Carlo results from a \supcell{3}{3}{3}
system are always higher in energy, presumably because the
true ground state can never fit into an odd dimensioned cell.)
Finally, we also exhaustively enumerated all $12^8$ orientation combinations
in the \supcell{2}{2}{1} super cell. The ground states found
in this way agree with the MC results whenever that shows a ground 
state with a \supcell{2}{2}{1} supercell, i.e. at strains less
than or equal to zero.  

\SAVE{Woosong sent lists of exact results, sorted by energies, 
from which we could can find the second and third best states, to
check if they are nearly degenerate with the best one.}

Brommer performed an 
exhaustive enumeration using a \supcell{\sqrt{2}}{\sqrt{2}}{2} supercell,
(this contains 8 clusters).
His best structure (see Ref.~\onlinecite{brommer-thesis}, Figure 4.13)
appears to be our a phase, containing the four orientations
$\dir{Z}{\pm}{r/l}$, which is similar but not identical to our
result at zero pressure, which was the ``$Z_2$'' phase.   
That is surprising, in view of the close similarity of our fitted potential
to that of Brommer {\it et al}
(see Sec.\ref{sec:cluster-int-brommer}, above).
\TODO{MM 12: What do you think of this disagreement?
Since Brommer fitted to a database of random clusters, whereas we
used the uniform background, I guess we can get slightly different
interactions.  Since the competing states seem to be very close
in energy, I guess it is possible to get a different answer.}

\subsubsection{Comparison with experiments}

\TODO{CLH: summarize the group-theoretical analysis of diffraction
by Tamura~\cite{tamura2005b}.}

\begin{figure}[h]
  \begin{center}
    \includegraphics[width=0.8\columnwidth]{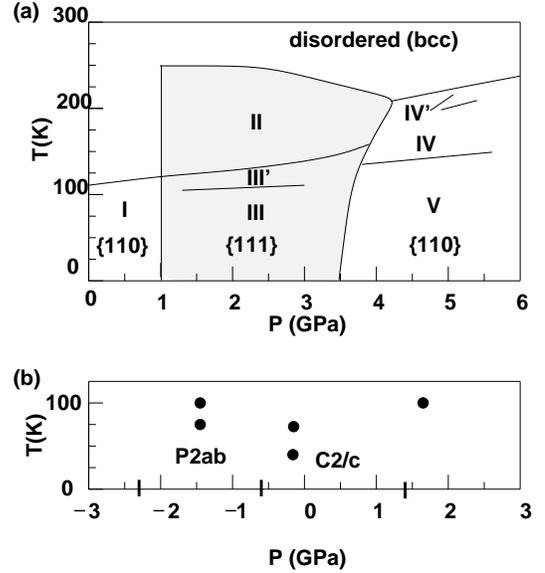}
  \end{center}
  \caption{(a) Pressure-temperature phase diagram of Cd$_6$Yb
[after Watanuki {\it et al}~\cite{watanuki2006}, Figure 3(a)].
The ordering wavevector of each phase is indicated (higher temperature
phases have similar order but partially disordered); region with $\{111\}$
phases is shaded. All transitions to the disordered (bcc) phase are
continuous.
(b) Our prediction for CaCd$_6$;  note shift in the pressure axes.
Critical pressures at $T=0$ are indicated by vertical bars; 
critical temperatures with first-order transitions, 
as reported in Table~\ref{tab:Tc-S} for the three strains where
this was measured, are shown by black circles.}
  \label{fig:P-T-diagram}
\end{figure}

Our results are reminiscent of, but not in agreement with, 
the experiments, which find ordered states with unit cells
of either \supcell{2}{2}{1} or \supcell{2}{2}{2}.
The experimental pressure-temperature ($P$--$T$) phase diagram
of Watanuki et al.~\cite{watanuki2006}
is shown in Figure~\ref{fig:P-T-diagram}.  
The ground state structure shown in Figure~\ref{fig:orderings}(b)
agrees with the $C2/c$ proposed arrangement in 
Ref.~\onlinecite{tamura2005a} similar to one in Ref.~\onlinecite{watanuki2006}

\LATER{
Compare with the ordering pattern with cubic symmetry in a 
\supcell{2}{2}{2} cell that is suggested by experiments,
in which each site has a $\langle 111 \rangle$ axis.
thus corresponding to three possible discrete orientations.
The simulation could be constrained to only allow one of those
three on each site (the three would be different on different
sites but we could call them X,Y,Z).}

\TODO{CLH 
How much do we know about the difference between 
Cd-Ca and Cd-Yb phase diagram/ordered states, anyhow?}

Direct comparison to experimental data is not possible 
since  the $P>0$ experiment is for a different alloy Cd$_6$Yb.
Still, some similarities exist.
Most importantly, the $Z_2$ phase, which we predict to be stable
in the pressure range bracketing $P=0$,
is (apart from small rotations around $z$) the same $C2/c$ structure
proposed in Figure 5 of Ref.~\onlinecite{tamura2005b}  for the 
(ambient pressure) phase of CaCd$_6$.
The $\{110\}$ structure suggested in Figure 3(c) of 
Ref.~\onlinecite{watanuki2006}, for the low- and high-pressure
phases of Cd$_6$Yb, is also the same as our $Z_2$ or $Z_4$ phase
if we use zero rotations in place of the small right/left rotations.

We have not studied our model system at the highest pressures, so
we do not know whether it has a counterpart for
the highest-pressure phase of Ref.~\onlinecite{watanuki2006}.

Watanuki {\it et al}~\cite{watanuki2006} 
suggest that the reason that different phases are
selected with increasing pressure is the competition of nearest-neighbor
and longer-range interatomic interactions. The latter were due, they 
conjectured, to Friedel oscillations. (We point out that elastically
mediated interactions are just as long-ranged as the oscillating
part of the interatomic potential, both of these being ideally $1/R^3$.)
However, we obtained a similarly
complex phase diagram without refitting the EAM potentials (meaning that
any change in Friedel oscillations was not explicitly taken into account).
Within our approach, the reason for the multiple phases is that the
orientational interaction of neighboring clusters is quite frustrated:
low-energy ordering patterns cannot simultaneously satisfy all interactions.
It seems there are several inequivalent ways to balance good interactions
with bad interactions that are nearly degenerate.  Thus, a small change
in the relative cost of different orientation combinations is expected 
to tip the balance to a different pattern.~\footnote{
A change in the relative strength 
of $\langle 1/2,1/2,1/2\rangle$ versus $\langle 1,0,0\rangle$ type
cluster pair interactions could be part of this change.  
This might be similar in its mechanism to the change posited in 
Ref.~\onlinecite{watanuki2006}, but is not the same, since they
imagined a change in the ratio of hard-core and long-range interactions.}

\subsection{Transitions at $T>0$}

Tamura {\it et al}~\cite{tamura2002} found an order-disorder transition
in CaCd$_6$at $T\approx 100$K. Under pressure, in the similar
system Cd$_6$Yb, Watanuki et al~\cite{watanuki2006} found order-disorder
transitions of the high-pressure phases too; order set in at
$T=$200 -- 250 K, and a further ordering within that phase occurred at
$T\approx 140$ K.

\begin{table}
\begin{tabular}{l|cccc}
\hline
strain  &  P    &  $T_c$  &  $\Delta E$    & $\Delta S$  \\
        & (GPa) &   (K)   & (meV/cluster)  &  ($k_B$/cluster) \\
\hline
$-0.010$ &$-1.46$& 72.5   &     3.359     &  0.537  \\
          &     &  97.5   &     6.953     &  0.827  \\
$0.000$  &$-0.14$& 40.0   &      0.547    & 0.159   \\
          &     &  70.0   &     3.672    &  0.609   \\
+0.010    & 1.65 & 100.0  &    6.641    &   0.770   \\
\hline
\end{tabular}
\caption{Thermal transitions at three fixed strains, along with
the corresponding pressure at $T=0$.
The $T_c$ is uncertain by $\pm 2.5$ K.  The jump in total energy
is shown for each transition, divided by the number of clusters,
and the corresponding entropy change $\Delta S$ per cluster 
is inferred (in units of Boltzmann's constant).
}
\label{tab:Tc-S}
\end{table}

Simulations by Brommer {\it et al} based on their version of the
discrete cluster pair Hamiltonian, using a \supcell{4}{4}{4} simulation
cell, found a first-order transition at $T\approx 91$K. Brommer pointed
out~\cite{brommer-thesis} this represents an entropy jump of about 1.0 
$k_B$ per cluster -- twice what was estimated by 
Tamura {\it et al}~\cite{tamura2003} for the similar compound Cd$_6$Y --
and an energy jumo of $\sim 10$meV/cluster.
\TODO{CLH look up that paper. 
Didn't Tamura later call it a second-order transition.}

At present we only have preliminary data concerning
transitions as a function of temperature.
The tempering MC simulation, which requires running multiple
replicas of the system at different temperatures,
naturally detects discontinuities in the energy as a 
function of temperature; our present results,
shown in Table~\ref{tab:Tc-S} , are based purely on this metric.  
We took data from zero to high temperatures for 
three different choices of the strain: strongly negative,
zero, and strongly positive, finding {\it two} first
order transitions for the first two cases, but only
one in the case of positive strains.  Around zero 
strain, there seems to
be a particular tendency to have closely competing states 
and low-energy excitations, and we believe this explains
the one rather low transition temperature for that case.

Simulations by Brommer {\it et al}~\cite{brommer2007} furthermore found 
a thermal ordering 
transition in a \supcell{4}{4}{4} supercell; they
did not analyze the orientation pattern in the ordered state there, 
but found its energy at the transition was only 1 meV/cluster higher than the
$\sqrt{2}\times\sqrt{2}\times 1$ structure. 
\SAVE{Ref.~\onlinecite{brommer-thesis}, Figure 4.13, shows a 
cell with only 4 clusters.  Thus I interpret it as the primitive
cell of the centered ``$Z_4$'' structure.}

Presumably, both in experiment and in our simulations,
the high-$T$  phases are partially disordered versions 
of the low-$T$ phases seen at the same pressure. 
In particular, the orientations related by
a change of $r$ to $l$ suffix in their symbols differ 
by a comparatively small rotation, so possibly 
(see Sec.~\ref{sec:justify-form}) they have similar interactions. 
Thus one possibility is that a partially disordered states
has a sublattice containing (say) random, equal populations
of $\dir{X}{+}{r}$ and $\dir{X}{+}{l}$ orientations.

\LATER{
The middle pressure phase of Ref.~\onlinecite{watanuki2006}
(in its high-temperature variant)  has a highly
symmetric space group $Fd\bar{3}$, in a 
$2\times 2 \times 2$ unit cell.
They proposed that, to realize this space group,
each tetrahedron should be oriented with the
maximum possible symmetry with respect to cubic
symmetry, i.e. with every corner pointing in
a $\langle 111 \rangle$ direction
[Figure 3(b) in Ref.~\onlinecite{watanuki2006}].
An argument against
their structure is that such orientations have a
high energy (as seen in the single-cluster effective
Hamiltonian).}

In many ternary cases, also in ScZn$_6$,
the ordering transition is not sharp in experiments, 
which Tamura speculated 
\LATER{Add citation}
indicates a glassy freezing.
This is plausible in view of the frustrated orientational interactions we found.

\section{Discussion}
\label{sec:discussion}

We will first summarize this work and then the outlook for
extensions of it.

\subsubsection{Summary}

In this paper, we presented a comprehensive template for 
studying cluster interactions in CaCd$_6$ and more generally
in any material possessing cluster orientation degrees of 
freedom.  We showed how to operationally relate cluster
orientations to atom positions (Sec.~\ref{sec:cluster+relax}),
worked out some group theory for the symmetry of the interactions
(Sections~\ref{sec:interaction-symmetries} and \ref{sec:cluster-pair-interaction}),
and fixed the technical obstacle of redundant parameters 
due to constraints in the pair counts (Sec.~\ref{sec:redundancies}).
Most of all, we showed that a singular value decomposition can
clarify the dominant nature of the interaction and allows a long 
list of fitting parameters to be truncated to a manageable number
(Sections~\ref{sec:SVD-method} and \ref{sec:results-SVD}).

When this was actually applied to CaCd$_6$, we found 
(in Sec.~\ref{sec:test-validity-discrete}
that the omitted interactions -- of whatever form, 
multi-cluster or pair interactions
with farther neighbors --  amounted to only 1/100-1/10 of the
nearest-neighbor pair interactions. The same is true
even for pair interaction contributions apart from
the first two singular vectors. This means that
truncating to those two dominant terms, 
as about as realisric as the nearest-neighbor Heisenberg spin exchange,
typically is, when used to model a magnetic material.

With effective interactions in hand, we carried out
Monte Carlo simulations of larger lattices, for the
purpose of discovering the optimum arrangements
(Section~\ref{sec:ordering}).
This is difficult, as the interactions are frustrated 
leading to glassiness.  Along with this, we made
a rather sketchy study of the phase diagram,
varying pressure and temperature.
Its overall nature is broadly reminiscent of experiment
(on CaCd$_6$ or isostructural compounds such as Cd$_6$Yb),
in that several ordering patterns are encountered
as pressure is varied, and there are also multiple
transitions with increasing temperature, which we 
respect represent partial orderings.

Using our method of orientation-constrained relaxation 
(Sec.~\ref{sec:constrained-relax}), it was possible to
extend our analysis to {\it continuously} varying  orientations
of the clusters.  For that case, augment couple the singular-value-decomposition
with a decomposition in terms of rotational harmonics.
We have not fully developed all that could be done with continuous
orientations. One byproduct of this calculation, which was not
available otherwise, is the {\it single-cluster} orientational
potential energy.  If the continuous type interactions can
be parametrized usefully for Monte Carlo simulations, one application
would be to study the {\it dynamics} of clusters including the paths
by which they pass from one discrete well to another.  (Such a
study was done for ScZn$_6$ using all-atom molecular dynamics with
pair potentials~\cite{mm2011-ScZn}.)

\subsubsection{Possible future work}

One obvious direction for future work is
to study the composition dependence.
For example, diffraction found somewhat different
ordering patterns as one varies
the large atom component in isostructural alloys
(Ca, Yb, and the other rare
earths have slightly different sizes):
e.g.  $\sqrt{2} \times  \sqrt{2} \times  2$ cell in 1/1-Cd$_6$Yb, 
a versus $\sqrt{2} \times \sqrt{2} \times 1$ C-centered monoclinic
cell in 1/1-Cd$_6$Ga.
Similarly, different critical temperatures
were measured experimentally.
Certainly, the strength of tetrahedron interactions 
will depend sensitively on the composition.
In ScZn$_6$~\cite{mm2011-ScZn}
the interactions are much weaker than in CaCd$_6$
but an ordering occurs nevertheless~\cite{Eu12}.
\SOON{Cite more on ScZn.}

Another direction is to thoroughly study the 
thermal behavior, which will require metrics
to identify the nature of partially ordered states.
This can also be extended to ``approximant''
phases with larger cells such as the $2/1$ approximant
of $i$-CdYb, which has four equivalent clusters per cubic cell;
this shows a transition to tetrahedron orientational
order, a complex stacking along a (100) direction~\cite{tamura2006}.

Our ultimate goal is to understand
the tetrahedron orientations in the {\it quasicrystal} phase,
and their role in stabilizing it.
A speculative possibility is the implementation of matching
rules by such clusters.
Matching rules in the Penrose tiling (or its 3D analog) 
are markings that spoil the symmetry of otherwise rhombic or pentagonal objects, 
and enforce a deterministic, quasi-periodic arrangement analogous
to an ideal crystal, thus offering one scenario for the stabilization
of quasicrystals.  It should be noted that the atomic structures of
various icosahedral quasicrystals, understood as packings of fully
symmetric clusters, showed no features that could implement such matching rules.
Furthermore, there is a more economical alternative scenario that
long-range order is emergent in dimension 3 from a ``random tiling'' 
in which the cluster packings sample an ensemble of extensive entropy.
The random-tiling scenario found support (in various icosahedral
quasicrystals, including $i$-CaCd) in the shapes of diffuse tails
around Bragg peaks in diffraction experiments 
\LATER{CITE de Boissieu papers}.
Otherwise no decisive experiments are known, so the 
best approach to discover matching rules, if they exist,
appears to be multi-scale simulation of the sort carried out
in the present paper.  The most plausible specific physical mechanism for 
matching rules in an icosahedral quasicrystal is via asymmetric
inner clusters such as the tetrahedron in $i$-CaCd. 
These might be investigated in the future by extending the 
methods of this study to larger approximants such as 
Ca$_{13}$Cd$_{76}$ \cite{gomez2001}.


\acknowledgments
This work was supported by
DOE Grant DE-FG02-89ER-45405 (WC, CLH, MM), and Slovak Research and Development Agency
funding under contracts VEGA 2/0111/11 and APVV-0076-11 (MM).
We are grateful to P. Brommer for providing the interaction
matrices computed in Refs.~\onlinecite{brommer2007} and 
\onlinecite{brommer-thesis}.

\appendix
\section{Comparison to Brommer {\it et al} fit}
\label{sec:cluster-int-brommer}

We were provided the interaction matrices computed
by Brommer {\it et al}~\cite{brommer2007}.
They had recognized the redundancies in the fit, and
resolved them arbitrarily by setting certain terms to zero. 
That meant the magnitude of any particular
pair interaction $V_{\alpha\beta}$ could not be given a physical interpretation:
such interactions are valid only for computing the total energy
of an entire configuration.
However, if one applies our recipe to resolve redundancies
from Sec.~\ref{sec:redundancies}
to their $V_{\alpha\beta}$, we get a well-defined result which 
can be compared with ours.

We expect a great similarity to our EAM results, since we
used the same potentials as they did.  
The important differences in our calculation are
\begin{itemize}
\item[(a)]
  For a database, out of the approaches we described in 
Sec.~\ref{sec:uniform-background},
they used random whole configurations
whereas we used the uniform background.
\item[(b)]
They did not use constrained relaxation, but relied on
the final states corresponding to the initial ones.
(See our discussion in Sec.~\ref{sec:constrained-relax}.)
Furthermore, their initial state was not one of the
twelve idealized orientations inferred from relaxations,
but one of the Gomez-Lidin orientations, postulated on
the basis of diffraction.
\TODO{CLH write down in more detail the extra symmetry.
I believe that Gomez and Lidin combined two partially
occupied sites from tetrahedron atoms.}
\SAVE{The results should be the same if the tetrahedra relax into
the same optimal orientations.}
\end{itemize}

\SAVE{Other differences from Brommer, that do not touch this comparison, were
our handling of the redundant degrees of freedom; and
using singular-value decomposition as a final stage in processing.
}

\LATER{CLH: Format and add the Brommer interaction matrices, which 
I have}

\begin{table}
 \begin{tabular}{c | c |   c c c c }
  \hline
 $\sigma$  & irrep & 
    $\dir{X}{+}{r}$& $\dir{X}{-}{l}$& $\dir{X}{+}{l}$& $\dir{X}{-}{r}$\\
  \hline
$27.69$ & E & $0.5542$ & $0.5542$ & $0.1617$ & $0.1617$  \\
        &   & $3.9$ & $-3.9$ & $-160.6$ & $160.6$ \\
$32.62$ & E & $0.3927$ & $0.3927$ & $0.4232$ & $0.4232$  \\
        &   & $-91.2$ & $91.2$ & $125.8$ & $-125.8$ \\
$22.6$  & E & $0.4139$ & $0.4139$ & $0.4025$ & $0.4025$ \\
        &   & $-104.3$ & $104.3$ & $-148.9$ & $148.9$ \\
$18.25$ & I & $0.3401$ & $-0.3401$ & $-0.2259$ & $0.2259$ \\
$13.15$ & I & $0.2887$ & $0.2887$ & $-0.2887$ & $-0.2887$ \\
$7.385$ & E & $0.1842$ & $0.1842$ & $0.5472$ & $0.5472$  \\
        &   & $44.8$ & $-44.8$ & $48.8$ & $-48.8$ \\
$5.404$ & I & $0.2259$ & $-0.2259$ & $0.3401$ & $-0.3401$ \\
  \hline
  \end{tabular}
  \caption{ Results from Brommer {\it et al}, Ref.~\onlinecite{brommer2007}:
singular values and right singular vectors for $c$-linkage,
same format as table~\ref{tab:cbondSVD}.  Note the symmetries
relating the pair of columns for $\dir{X}{+}{r}$ and $\dir{X}{-}{r}$,
and similarly relating the columns for $\dir{X}{+}{l}$ and $\dir{X}{-}{l}$:
they have the same magnitudes and, for representation $E$, they have
opposite phase angles (after removal of an overall arbitrary phase).
  \TODO{CLH+ Since Brommer's interaction table was so similar to ours,
why do the c-bond irreps look so different?  
We might think,  the Brommer tetrahedron orientations are
different from ours, and when we try to implement a 3-fold symmetry, could that 
screw it up?  Or could Woosong's convention for $r/l$ with $-$ orientations?
I don't think either matters. 
The singular values themselves are computed from the raw interaction matrix,
so they should be immune to any labeling confusion.
I am worried that, e.g., the wrong four columns were picked out.  }
}
\label{tab:cbondSVD-brommer}
\end{table}

The results look similar for the $b$-linkage interaction  matrix.
When we perform the SVD analysis, the overall pattern of singular
values seen in Table~\ref{tab:bbondSVD-brommer}
is similar to what we saw in Table~\ref{tab:bbondSVD}),
though with slightly less of a separation between large and small
singular values. However, the dominant singular vector comes from
a different representation, and the pattern of signs in the
singular vectors is also different.  (It is likely these differences 
are artifacts of different conventions we used for the orientations;
Ref.~\cite{brommer2007} used the other choice for orienting icosahedral
axes in cubic symmetry, differing from ours by a 90$^\circ$ rotation.

On the other hand, the results for the $c$-linkage interaction matrix
look completely different, as does the singular value decomposition
(Table ~\ref{tab:cbondSVD-brommer}).
Furthermore, for reasons we do not understand, the latter
exhibits an additional symmetry relating the  columns
differing by the sense $+$ and $-$ of the orientation,
which does not correspond to {\it any} symmetry of
the clusters.  Finally, the $c$-linkage singular values
are large and (unlike all other cases analyzed in this
paper) the magnitudes of the largest and smallest ones 
do not differ by a large factor.

\CLH{(WC sent CLH the table for (001) linkage: CLH put in the
file ``woosong-brommer-b-0808'')}
\NOTE{It does look quite nice, except I still have to figure out the coordinate 
notation. It seems to have swapped blocks per my naive understanding. Maybe different
handedness [of the setting. See next note to MM]}
\TODO{CLH. Incorporate into text 
Brommer's b-linkage and c-linkage interactions, and 
SVD analysis of them. (Put by WC as comment in text 9/11/12 draft, 
saved under ``prog/qx/woosong/Results/E-pair/b+c-int+SVD\_brommer''}

\TODO{CLH? After understand the orientations, convert table to same format
used in the paper.}

\TODO{CLH: (related to the just-preceding notes) We discussed that
two inconsistent settings of the framework (differing by 4-fold 
rotation) are present in MM data set.  Those used by Brommer et al,
seem to differ from those which MM provided to WC?  We see a similar
pattern in the singular vectors, but with a switch of the
coordinates?
}

\begin{table}
  \centering
  \begin{tabular}{c | c | c c c c }
\hline
$\sigma$ (meV)  & irrep & $\dir{X}{+}{r}$&$\dir{Y}{+}{r}$&$\dir{Z}{+}{r}$&$\dir{Z}{-}{l}$\\
    \hline  
\hline
69.37  & $B_2$ & $-0.1587$&    $-0.4741$&     0   &        0  \\
21.75  & $A_1$ & $-0.0851$&    0.1515  &   0.3978 &   $-0.5305$ \\
6.194  & $A_1$ & 0.0064  &     0.3327  &  $-0.4950$&  $-0.1832$ \\
5.415  & $B_2$ & $-0.4741$&    0.1587  &      0   &        0  \\
4.787  & $B_1$ & 0.4921  &     0.0885  &      0   &        0  \\
4.280  & $A_2$ & 0.1681  &     0.1638  &   0.4602 &   $-0.4220$ \\
2.415  & $A_2$ & 0.0087  &     $-0.4041$&  0.3967 &     0.1259  \\
1.672  & $A_2$ & 0.1684  &     $-0.2417$& $-0.3588$&  $-0.4446$ \\
8.392$\times 10^{-1}$ &  $B_1$ &
  0.0885  &  $-0.4921$    &      0 &    0     \\
7.456$\times 10^{-1}$ &  $A_2$ &
 $-0.4396$ &   $-0.0380$  &   0.0464  &  $-0.3292$  \\
7.442$\times 10^{-2}$ &  $A_1$ &
  0.3992  &   $-0.1818$ &   $-0.1160$ &  $-0.3189$  \\
\hline
  \end{tabular}
\caption{ Results from Brommer {\it et al}, Ref.~\onlinecite{brommer2007}:
singular values and right singular vectors for $b$-linkage,
same format as table~\ref{tab:bbondSVD}. }
\label{tab:bbondSVD-brommer}
  \end{table}

\end{document}